\documentclass[a4paper,11pt]{article}
\pdfoutput=1
\usepackage{jheppub}
\pdfoutput=1
\usepackage{epsfig,graphicx}
\usepackage[english]{babel} 
\usepackage{xcolor}
\definecolor{darkgreen}{HTML}{4CA64C} 
\usepackage{pifont}
\usepackage[utf8]{inputenc} 
\usepackage{amssymb} 
\usepackage[export]{adjustbox}
\usepackage{amsmath} 
\usepackage{xspace} 
\usepackage{esint}
\usepackage{xcolor}
\usepackage{amssymb}
\usepackage{afterpage}
\usepackage[toc,page]{appendix}
\usepackage{url} 
\usepackage{braket} 
\usepackage{slashed} 
\usepackage{multirow} 
\usepackage{bbold} 
\usepackage{dsfont} 
\usepackage{mathrsfs}
\usepackage[dvipsnames,svgnames,table]{xcolor}  
\usepackage{marginnote} 
\usepackage{minitoc} 
\usepackage{epigraph} 
\usepackage{comment}
\usepackage{pdfpages} 
\usepackage{float}
\usepackage{chngcntr}
\usepackage{fancyhdr}
\makeatletter
\renewcommand\@mkboth[2]{\markboth{#1}{}}
\makeatother
\usepackage{epsf}
\usepackage{epsfig}
\usepackage{subfig}
\usepackage{mathtools}    
\usepackage{float}
\usepackage{multirow}
\usepackage{nicefrac}
\usepackage{epstopdf}
\usepackage{slashed}
\usepackage{nameref}
\usepackage{xcolor}
\usepackage{url}
\usepackage{breqn}
\usepackage{amsmath}   
\usepackage{adjustbox} 
\usepackage{chemformula}
\usepackage{multicol}
\usepackage{placeins}
\usepackage{braket}
\usepackage{subdepth}
\usepackage{color}
\usepackage{booktabs}
\usepackage{siunitx} 
\allowdisplaybreaks

\DeclareMathOperator{\Li}{Li}

\def\nbslash{\rlap{\hspace{0.02cm}/}{\bar n}}
\def\nslash{\rlap{\hspace{0.02cm}/}{n}}
\newcommand{\eps}{\varepsilon}
\newcommand{\IR}{\scriptscriptstyle \mathrm{IR}}
\newcommand{\UV}{\scriptscriptstyle \mathrm{UV}}
\usepackage[sort&compress,numbers]{natbib}
\usepackage{doi}
\usepackage{hyperref} 
\usepackage[capitalise]{cleveref}
\usepackage[dvipsnames,svgnames,table]{xcolor} 
\def\nbslash{\rlap{\hspace{0.02cm}/}{\bar n}}
\def\nslash{\rlap{\hspace{0.02cm}/}{n}}
\graphicspath{ {./Figures/} }

\title{Resolved $Q_1^q$-$Q_{7\gamma}$ interference in $\bar{B}\to X_s\gamma$ at NLO: Hard matching  and radiative charm-loop jet function for $m_c \to 0$}

\author{Riccardo Bartocci$^{\,a,b}$, Philipp B\"oer$^{\,c}$, Kevin Brune$^{\,d}$ and Tobias Hurth$^{\,d}$} 

\emailAdd{riccardo.bartocci@kit.edu}
\emailAdd{philipp.boeer@cern.ch}
\emailAdd{kbrune@uni-mainz.de}
\emailAdd{hurth@uni-mainz.de}

\affiliation{${}^a$Institute for Theoretical Physics, Karlsruhe Institute of Technology, Wolfgang-Gaede-Str. 1, 76131 Karlsruhe,
Germany}
\affiliation{${}^b$Institute for Astroparticle Physics, Karlsruhe Institute of Technology, 76344 Eggenstein-Leopoldshafen,
Germany}
\affiliation{${}^c$CERN, Theoretical Physics Department, CH-1211 Geneva 23, Switzerland}
\affiliation{${}^d$PRISMA+ Cluster of Excellence \& Institute of  Physics (THEP) \& Mainz Institute for Theoretical Physics, Johannes Gutenberg University, D-55099 Mainz, Germany}

\abstract{
We study next-to-leading-order (NLO) QCD corrections to the subleading-power resolved-photon contribution to $\bar{B}\to X_s\gamma$ arising from the interference of the current-current operator $Q_1^q$ and the electromagnetic dipole operator $Q_{7\gamma}$. This contribution is among the main sources of theoretical uncertainty in Standard Model predictions for the decay rate. Using a factorisation theorem in soft-collinear effective theory for the kinematic endpoint region, $m_b-2E_\gamma=\mathcal{O}(\Lambda_{\rm QCD})$, we determine all previously unknown perturbative ingredients at NLO in the limit of a vanishing charm-quark mass, $m_c\to0$. Specifically, we compute the $m_c$-independent one-loop hard-matching coefficients, derive their renormalisation-group equations, and obtain accurate analytic approximations to their solutions by retaining the first two terms in a $1/N_c$ expansion. Our second main result is the two-loop correction to the radiative charm-loop jet function, including its full dependence on the collinear momentum fraction. Combining these results with the known $\mathcal{O}(\alpha_s)$ correction to the quark jet function and the one-loop renormalisation of the subleading shape function $g_{17}$, previously computed by some of us, we provide the first complete NLO description of this resolved contribution in the massless-charm limit. This leaves the $m_c$ dependence of the radiative charm-loop jet function as the remaining ingredient for a phenomenological analysis within renormalisation-group-improved perturbation theory.
}

\preprint{
\begin{minipage}{4cm}
\small
  \flushright
MITP/26-061
CERN-TH-2026-243
P3H-26-092
KA-TP-32-2026
\end{minipage}}

\begin{document}

\maketitle

\section{Introduction} 
\label{sec:introduction}
The inclusive radiative decay $\bar{B}\to X_s\gamma$ provides a precision test of the quark flavour sector of the Standard Model (SM) and a sensitive indirect probe of new physics.  Power-suppressed resolved contributions are still one of the largest theoretical uncertainties of the CP- and isospin-averaged branching ratio of the $\bar{B}\rightarrow X_s \gamma$ decay~\cite{Benzke:2020htm,Misiak:2026sqy}. Here, we focus on the resolved contribution induced by the interference of the current-current operators $Q_1^q$ with the electromagnetic dipole operator $Q_{7\gamma}$,
\begin{equation}
    \label{eq:Q1Q7-definition}
    Q_1^q = [\bar{q}_i\gamma^{\mu}(1-\gamma_5)b_i][\bar{s}_j\gamma_{\mu}(1-\gamma_5)q_j] \,, \qquad Q_{7\gamma}=\frac{-em_b}{8\pi^2}\bar{s}\sigma_{\mu\nu}(1+\gamma_5)F^{\mu\nu}b,
\end{equation}
where $q=u,c$ and $i,j$ are colour indices. Operator mixing under renormalisation also requires the inclusion of
\begin{equation}
    \label{eq:Q2-definition}
    Q_2^q = [\bar{q}_i\gamma^{\mu}(1-\gamma_5)b_j][\bar{s}_j\gamma_{\mu}(1-\gamma_5)q_i] \,.
\end{equation}
The substantial scale ambiguity and charm-mass dependence of this resolved-photon contribution at LO motivate a systematic calculation of radiative corrections within renormalisation-group (RG) improved perturbation theory~\cite{Gunawardana:2019gep, Benzke:2020htm}, which we recently initiated~\cite{Bartocci:2024bbf,Bartocci:2025lbl,Bartocci:2026eeq} and continue in this work.

We consider the photon-energy spectrum in the endpoint region, $m_b-2E_\gamma=\mathcal{O}(\Lambda_{\rm QCD})$, where soft-collinear effective theory (SCET)~\cite{Bauer:2000ew,Bauer:2000yr,Beneke:2002ph} separates perturbative and non-perturbative
dynamics.
The corresponding factorisation formula was established at LO in~\cite{Benzke:2010js}. A generalisation to higher-order corrections was recently presented in~\cite{Bartocci:2026eeq}.
Schematically, it reads
\begin{equation}
    \label{eq:factorisationtheorem}
    \frac{\mathrm{d}\Gamma_{17}(\bar{B}\rightarrow X_s \gamma)}{\mathrm{d} E_\gamma} \sim \, H^*_{7\gamma}\, \int_0^1 \mathrm{d}u\, H_1^q(u) \int_{-\infty}^{\infty} \mathrm{d}\omega_1\, \bar{J}(u,\omega_1)\int_{-\infty}^{\bar{\Lambda}} \mathrm{d}\omega\, J(\omega)\, g_{17}(\omega,\omega_1) \,,
\end{equation} 
with $\bar{\Lambda} = m_B - m_b$. Here, $u$ denotes the collinear momentum fraction of the quark emitted from the four-quark operator relative to the photon, while $\omega$ and $\omega_1$ describe the soft momenta of the heavy-quark field and the gluon-field of the $\bar{B}$ meson, respectively. The hard functions $H_{7\gamma}$ and $H_1^q(u)$ combine the Wilson coefficients of the weak effective Hamiltonian with the hard-matching coefficients onto SCET. The radiative charm-loop jet function $\bar{J}(u,\omega_1)$ describes the perturbative dynamics associated with the conversion of a quark--antiquark pair from the current-current operator into an energetic photon through an interaction with the soft gluon field of the $\bar{B}$ meson. The hardcollinear quark jet function $J$, which describes the hadronic final state, is known through N$^3$LO~\cite{Bosch:2004th,Becher:2006qw,Bruser:2018rad}. Non-perturbative dynamics is encoded in the subleading shape function $g_{17}(\omega,\omega_1)$, whose one-loop RG equation was derived and solved by some of us in~\cite{Bartocci:2024bbf}.

The objective of the present work is twofold. First, we determine the one-loop hard matching coefficients and derive their evolution equations in the factorisation scale. For $H_1^q(u)$, we provide analytic solutions
for the first two terms in the large-$N_c$ expansion. Second, we compute the two-loop radiative charm-loop jet function $\bar J(u,\omega_1)$ in the massless-charm limit, extending the earlier integrated result~\cite{Bartocci:2026eeq} to retain the full dependence on the momentum fraction $u$. We also derive its one-loop renormalisation factor and verify that it correctly predicts the singularities of our two-loop calculation. This provides an independent consistency check and constrains the corresponding
calculation for finite $m_c$. Together with the known one-loop corrections to the quark jet
function~\cite{Bosch:2004th} and the one-loop evolution kernel of the subleading shape function $g_{17}$~\cite{Bartocci:2024bbf}, our results complete the perturbative framework for an RG-improved treatment of this contribution at NLO in the limit $m_c\to0$. This establishes a complete example of NLO factorisation and RG improvement beyond leading power for an inclusive observable. The charm-mass dependence of the two-loop radiative jet function remains the only missing perturbative ingredient for phenomenological applications, and our analytic results provide a benchmark for its calculation.

The remainder of this article is organised as follows. In Section~\ref{sec:hardmatching}, we determine the hard functions by matching the weak effective operators onto SCET and derive the corresponding anomalous dimensions. In Section~\ref{sec:RGEsolution}, we discuss the RG evolution of the hard functions and obtain an approximate analytic solution to the non-local evolution equation using a large-$N_c$ expansion. In Section~\ref{sec:antijet_function}, we compute the radiative charm-loop jet function at LO and NLO in the limit $m_c\to0$. We conclude in Section~\ref{sec:conclusion}. Several appendices summarise known results and provide further technical details of the calculations.

\section{Matching to SCET}
\label{sec:hardmatching}
Perturbative QCD corrections from the hard scale $\mu \sim m_b$ are included by matching the electroweak operators $Q_{7\gamma}$ and $Q_{1,2}^q$ to SCET. The corresponding matching coefficients enter the hard function in the factorisation formula~\eqref{eq:factorisationtheorem} for the differential decay width. The purpose of this section is to provide the renormalisation factors of the effective operators, the UV- and IR-finite hard corrections and their evolution equations that resum large logarithms between the hard and hardcollinear scales. 

\subsection{Matching of \texorpdfstring{$Q_{7\gamma}$}{Q7γ}}
\label{subsec:hardmatching_Q7}

The electromagnetic dipole operator $Q_{7\gamma}$ is matched to the SCET operator
\begin{equation}
   {\cal O}_{7 \gamma} = \frac{-e \overline{m}_b(\nu)}{4\pi^2} \, \bar{\chi}_{hc} \frac{\slashed{\bar{n}}}{2} [in\cdot\partial \slashed{A}_{\perp,\overline{hc}}^{\rm em}](1+\gamma_5)h_v 
\end{equation}
according to $Q_{7\gamma}(\nu) = H_{77}(\nu,\mu) \, {\cal O}_{7\gamma}(\nu,\mu)$, with the scale $\nu$ of the Wilson coefficient and $\mu$ the factorisation scale. Here, $\bar{\chi}_{hc} = \bar{\xi}_{hc} W_{hc}$ denotes the gauge-invariant hardcollinear quark building block, where $\xi_{hc}$ is the hardcollinear quark field and $W_{hc}$ the corresponding Wilson line. The transverse photon field is denoted by $A_{\perp,\overline{hc}}^{\text{em}, \mu}$. The prefactor $\overline{m}_b(\nu)$ is the renormalised $b$-quark mass in the $\overline{\mathrm{MS}}$ scheme.

\paragraph{SCET renormalisation factor:} To subtract the infrared (IR) divergences from the hard-matching coefficient, we first compute the SCET renormalisation factor defined by $\mathcal{O}_{7\gamma, \text{ren}} = Z_{\mathcal{O}_{7\gamma}} \mathcal{O}_{7\gamma,  \text{bare}}$ in the $\overline{\mathrm{MS}}$ scheme. Since the $b$-quark mass is renormalised separately at the scale $\nu$ and treated as a fixed parameter under the SCET renormalisation at the factorisation scale $\mu$, the renormalisation factor $Z_{\mathcal{O}_{7\gamma}}$ accounts only for the ultraviolet divergences of the SCET operator and does not include the QCD mass renormalisation factor $Z_{m_b}$, cf.~Appendix~\ref{app:evanescent_ops}.

\begin{figure}[t]
  \centering
  \begin{minipage}[t]{0.49\textwidth}
    \centering
    \includegraphics[width=0.45\textwidth]{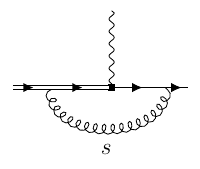}
    \caption{Soft one-loop diagram contributing to the renormalisation of the operator $\mathcal{O}_{7\gamma}$. The double line denotes the heavy-quark field in HQET. Field-renormalisation diagrams are not shown.}
    \label{fig:O7_SCET_ren_soft}
  \end{minipage}
  \hfill
  \begin{minipage}[t]{0.49\textwidth}
    \centering
    \includegraphics[width=0.45\textwidth]{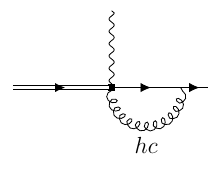}
    \caption{Hardcollinear one-loop diagram contributing to the renormalisation of the operator $\mathcal{O}_{7\gamma}$. Field-renormalisation diagrams are not shown.}
    \label{fig:O7_SCET_ren_hc}
  \end{minipage}
\end{figure}

We work in Feynman gauge. The ultraviolet (UV) divergences of the soft and hardcollinear loops shown in Figs.~\ref{fig:O7_SCET_ren_soft} and~\ref{fig:O7_SCET_ren_hc} are extracted by evaluating the integrals in $D = 4 - 2\eps$ dimensions while taking the external collinear momentum $p$ of the energetic strange quark off shell, with $p^2 \equiv p_+ p_- + i\eta$ and $p_- = m_b$ the pole mass. In this regularisation scheme, the soft loops yield
\begin{align}
\label{eq:ZH}
 \delta Z_{\mathcal{H}}(\mu) = Z_{\mathcal{H}}(\mu) - 1 =  - \frac{\alpha_s C_F}{4\pi} \left(- \frac{1}{\eps^2} + \frac{2}{\eps} \ln \frac{-p^2}{m_b \mu} + \frac{1}{\eps} \right) + \mathcal{O}(\alpha_s^2) \,, 
\end{align}
which corresponds to the UV poles of the gauge-invariant heavy-quark field $\mathcal{H}_v \equiv S_n^\dagger h_v$, i.e.\ the static $h_v$ field dressed with the soft light-like Wilson line along the direction of the energetic strange quark. The hardcollinear loops yield the known result for the off-shell regulated gauge-invariant quark field~\cite{Beneke:2020vnb}
\begin{align}
\label{eq:Zchi}
   \delta Z_\chi(\mu) = Z_\chi(\mu) - 1 = - \frac{\alpha_s C_F}{4\pi} \left( \frac{2}{\eps^2} - \frac{2}{\eps} \ln \frac{-p^2}{\mu^2} + \frac{3}{2\eps} \right) + \mathcal{O}(\alpha_s^2) \,. 
\end{align}
The sum of the two contributions determines the one-loop renormalisation factor, 
\begin{align}
\label{eq:Z7gamma}
   Z_{\mathcal{O}_{7\gamma}}(\mu) = 1 + \delta Z_{\mathcal{H}}(\mu) + \delta Z_\chi(\mu) + \mathcal{O}(\alpha_s^2) 
   = 1 - \frac{\alpha_s C_F}{4\pi} \left( \frac{1}{\eps^2} + \frac{1}{\eps} \ln \frac{\mu^2}{m_b^2} + \frac{5}{2\eps} \right) + \mathcal{O}(\alpha_s^2) \,, 
\end{align}
which is independent of the off-shellness $p^2$ as expected.

\begin{figure}[t]
    \centering
     \includegraphics[width=0.3\textwidth]{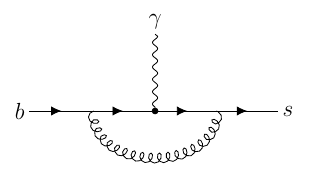}
    \caption{One-loop hard-matching correction contributing to $H_{77}$. Again, field-renormalisation diagrams are not shown.}
    \label{fig:hard_7}
\end{figure}

\paragraph{Hard matching coefficient:} The hard-matching coefficient $H_{77}$ can be extracted from the $b \to s \gamma$ amplitude with on-shell partons. At NLO, the matrix element of the bare dipole operator $Q_{7\gamma}$, expressed in terms of renormalised quark fields, is obtained by combining the contribution from the diagram in Fig.~\ref{fig:hard_7} with the on-shell wave-function
renormalisation contributions of the external quarks given in Appendix~\ref{app:evanescent_ops}. We find
\begin{align}
\label{eq:H7poles}
\langle Q_{7\gamma}^{\text{bare}}\rangle^{(1)}
= -\frac{\alpha_s C_F}{4\pi}
\Bigg(
&\frac{1}{\eps_{\IR}^2}
+ \frac{5}{2\eps_{\IR}}
+ \frac{1}{\eps_{\IR}} \ln\frac{\mu^2}{m_b^2}
+ \frac{1}{\eps_{\UV}}  \nonumber \\
+ &\frac{1}{2}\ln^2\frac{\mu^2}{m_b^2}
+ \frac{5}{2}\ln\frac{\mu^2}{m_b^2}
+ \ln\frac{\nu^2}{m_b^2}
+ 6
+ \frac{\pi^2}{12}
\Bigg)
\langle \mathcal{O}_{7\gamma}\rangle^{(0)} \,,
\end{align}
where we distinguish ultraviolet (UV) and infrared (IR)
poles. Since the vertex correction itself is UV finite
in Feynman gauge, the UV pole in~\eqref{eq:H7poles}
originates entirely from the field renormalisation contributions. It is cancelled by the $b$-quark mass-renormalisation factor and the renormalisation of the composite operator $Q_{7\gamma}$
(see Appendix~\ref{app:evanescent_ops}). The remaining IR singularities match the UV pole structure of the corresponding SCET operator in~\eqref{eq:Z7gamma}. We can then read off the diagonal part $H_{77}$ of the UV- and IR-finite hard matching coefficient,
\begin{align}
\label{eq:H77}
   H_{77}(\mu,\nu) &= 1 - \frac{\alpha_s(\nu) C_F}{4\pi} \left( \frac12 \ln^2\frac{\mu^2}{m_b^2}
    + \frac52 \ln\frac{\mu^2}{m_b^2} + \ln\frac{\nu^2}{m_b^2} + 6 + \frac{\pi^2}{12} \right) \,.
\end{align}

At $\mathcal{O}(\alpha_s)$, the weak operators $Q_1^q$ and $Q_{8g}$ also contribute to the matching onto $\mathcal{O}_{7\gamma}$, with IR-finite matching coefficients~\cite{Neubert:2004dd},
\begin{align}
\label{eq:H17}
   H_{17}^q(\nu) &= \frac{\alpha_s(\nu) C_F}{4\pi}
    \left( - \frac{52}{27} \ln\frac{\nu^2}{m_b^2} - \frac{833}{162} - \frac{20\pi i}{27} \right) , \\[0.2em]
\label{eq:H87}
   H_{87}(\nu) &= \frac{\alpha_s(\nu) C_F}{4\pi}
    \left( \frac43 \ln\frac{\nu^2}{m_b^2} + \frac{11}{3} - \frac{2\pi^2}{9} + \frac{2\pi i}{3} \right) .
\end{align}
Consistent with the adopted power counting $m_c^2=\mathcal{O}(m_b\Lambda_{\rm QCD})$, we evaluate the hard matching coefficients in the limit $z\equiv m_c^2/m_b^2\to0$. The constant in $H_{17}$ then corresponds to the massless limit $g(z=0) = -833/162 - 20 \pi i/27$ of the penguin function given in~\cite{Neubert:2004dd}. It is obtained from the renormalised two-loop $b \to s\gamma$ matrix element of the current-current operator~\cite{Greub:1996tg}. Numerically relevant finite-charm-mass terms could be included as power corrections~\cite{Greub:1996tg,Neubert:2004dd}.

Lastly, the hard function entering the factorisation formula is obtained by including the respective Wilson coefficients of the weak effective Hamiltonian,
\begin{equation}
\label{Hgamma}
   H_{7\gamma}(\nu, \mu) = C_1^q(\nu)\, H_{17}^q(\nu) + C_{7\gamma}(\nu)\, H_{77}(\mu,\nu) + C_{8g}(\nu)\, H_{87}(\nu) \,. 
\end{equation}
We note that only the product $\overline{m}_b(\nu) H_{7\gamma}(\nu, \mu)$ is independent of the scale $\nu$, and that employing the effective coefficients $C_{7\gamma}^{\mathrm{eff}}
= C_{7\gamma} - C_5/3 - C_6$ and $C_{8g}^{\mathrm{eff}} = C_{8g} + C_5$ accounts for the leading contributions from the four-quark penguin operators $Q_5$ and $Q_6$.
Additional penguin contributions from the four-quark
operators can be neglected, as they are numerically small
\cite{Neubert:2004dd}.
\paragraph{Anomalous dimension:}
The renormalisation factor~\eqref{eq:Z7gamma} determines the evolution of the SCET operator with the factorisation scale,
\begin{equation}
    \frac{d}{d\ln\mu}\, \mathcal{O}_{7\gamma}(\mu) = -\gamma_J(\mu)\, \mathcal{O}_{7\gamma}(\mu) \,,
    \qquad
    \gamma_J = -\frac{dZ_{\mathcal{O}_{7\gamma}}}{d\ln\mu}\, Z_{\mathcal{O}_{7\gamma}}^{-1} \,.
\end{equation}
Since the product of the hard function and the SCET operator is independent of $\mu$, the hard function obeys
\begin{equation}
    \label{eq:rge_H7}
    \frac{d}{d\ln\mu}\, H_{7\gamma}(\mu,\nu) = \gamma_J(\mu)\, H_{7\gamma}(\mu,\nu) \,.
\end{equation}
Following the notation of~\cite{Neubert:2004dd,Becher:2006nr},
we write the anomalous dimension of the heavy-to-light
current~\cite{Bauer:2000ew,Bauer:2000yr} as
\begin{equation}
\label{eq:Gamma7}
    \gamma_J(\mu) = - C_F\, \gamma_{\rm cusp}(\alpha_s) \ln\frac{\mu}{m_b} + \gamma'(\alpha_s)
    = - \frac{\alpha_s C_F}{4\pi} \left( 4 \ln\frac{\mu}{m_b} + 5 \right) + \mathcal{O}(\alpha_s^2) \,,
\end{equation}
where $\gamma_{\rm cusp}(\alpha_s) = \sum_{n\geq 0} \gamma^{\rm cusp}_n \left(\frac{\alpha_s}{4\pi}\right)^{n+1}$ with $\gamma^{\rm cusp}_0 = 4$ is the cusp anomalous dimension with the colour factor stripped off, so that $\Gamma_{\rm cusp} = C_F\, \gamma_{\rm cusp}$ for a quark, and $\gamma'(\alpha_s) = \sum_{n\geq 0} \gamma'_n \left(\frac{\alpha_s}{4\pi}\right)^{n+1}$ with $\gamma'_0 = -5 C_F$. Equation~\eqref{eq:Gamma7} coincides with the anomalous dimension of the heavy-to-light current at $2E_\gamma = m_b$ given in~\cite{Neubert:2004dd}. 
Moreover, for $\mu = \nu$ the hard function~\eqref{Hgamma} agrees with the result of~\cite{Neubert:2004dd}. Identifying these scales may also be used for RG evolution, which allows one to evolve all functions to the hardcollinear scale, including the Wilson coefficients and the $b$-quark mass. Recall that the dependence on the scale $\nu$ cancels in the product $\overline{m}_b(\nu) H_{7\gamma}(\nu, \mu)$. Thus, the only non-trivial running in this product arises from the factorisation scale $\mu$.

\subsection{Matching of \texorpdfstring{$Q_{1,2}^q$}{Q1,2^q}}
\label{subsec:hardmatching_Q12}
The current-current operators $Q_{1,2}^q$ are matched to the SCET operators
\begin{align}
\label{eq:SCETO12}
 {\cal O}_1^{q}(s) &= [\bar{\chi}_{hc}^j(0)   \gamma_{\perp\mu}(1-\gamma_5) \chi_{\overline{hc}}^{j}(0)] \,  [\bar{\chi}_{\overline{hc}}^{i}(sn) \gamma^\mu_\perp (1-\gamma_5) h_v^i(0)] \,,  \nonumber \\[0.2em]
 {\cal O}_2^{q}(s) &= [\bar{\chi}_{hc}^j(0)   \gamma_{\perp\mu}(1-\gamma_5)  \chi_{\overline{hc}}^{i}(0)] \, [\bar{\chi}_{\overline{hc}}^{i}(sn) \gamma^\mu_\perp (1-\gamma_5) h_v^j(0)] \,, 
\end{align}
which contain two anti-hardcollinear quark fields that are separated on the light cone. This nonlocality reflects that hard loops redistribute the large photon energy between the quark and antiquark, which carry momentum fractions $u$ and $\bar{u}=1-u$, respectively. Consequently, the momentum-space hard-matching coefficients become functions of $u$. Radiative corrections mix the two colour structures in~\eqref{eq:SCETO12}, such that we define a matrix of hard-matching coefficients as
\begin{equation}
\label{eq:H12def}
    Q_i^q(\nu) = \sum_{n=1,2} \int_0^1 \! du \, \big[\mathbf{H}_{12}(u;\mu,\nu)\big]_{in} \,  \tilde{\mathcal{O}}_n^q(u;\mu) \,, 
\end{equation}
where $\tilde{\mathcal{O}}_n^q(u;\mu)$ denotes the Fourier transform of~\eqref{eq:SCETO12} in the light-cone separation $s$. We note that, in addition, evanescent operators (which vanish in $D=4$ but are non-zero in $D \neq 4$ space-time dimensions) are required to close the operator basis under renormalisation. They give rise to finite contributions to the hard-matching coefficients, which we discuss in detail in Appendix~\ref{app:evanescent_ops}. 

\paragraph{SCET renormalisation factor:}
As before, to subtract the IR divergences from the matching coefficients, we first compute the now matrix-valued renormalisation factor $\mathbf{Z}_{\mathcal{O}_{12}}$ of the effective operators,
\begin{equation}
    \tilde{\mathcal{O}}_{\text{ren}, \, n}^q(u;\mu) = \int_0^1 \! dv \, (\mathbf{Z}_{\mathcal{O}_{12}})_{nm}(u,v;\mu) \, \tilde{\mathcal{O}}^q_{{\rm bare}, \, m}(v;\mu) \,, 
\end{equation}
with $n,m = 1,2$. Here, we write $\mathbf{Z}_{\mathcal{O}_{12}}(u,v;\mu) = \mathbb{1}_{2\times 2}\,\delta(u-v) + \delta\mathbf{Z}_{\mathcal{O}_{12}}(u,v;\mu)$ and determine the NLO correction from the soft, hardcollinear and anti-hardcollinear loops separately.

\begin{figure}[t]
  \centering
  \includegraphics[width=0.65\textwidth]{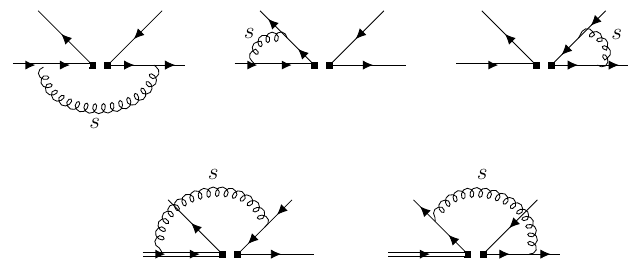}
  \caption{Soft one-loop diagrams that contribute to the renormalisation of the operators $\mathcal{O}_{1,2}^q$. Here, the horizontal lines represent the incoming $b$ quark and the outgoing collinear $s$ quark, whereas the remaining two lines represent the outgoing anti-hardcollinear quark-antiquark pair.}
  \label{fig:O1_SCET_ren_soft}
\end{figure}

Since soft exchanges do not change the large momentum component $u$ of the anti-hardcollinear quarks, their contribution to the renormalisation factor is local,
\begin{equation}
    \delta\mathbf{Z}_{\mathcal{O}_{12}}^{\rm soft}(u,v;\mu) = 
    \delta(u-v) \, \delta\mathbf{z}_{\rm soft}(\mu) \,. 
\end{equation}
By evaluating the diagrams in Fig.~\ref{fig:O1_SCET_ren_soft}, one finds that the matrix $\delta\mathbf{z}_{\rm soft}$ takes the structure
\begin{equation}
\label{eq:deltazsoft}
 \delta\mathbf{z}_{\rm soft}(\mu) = \delta Z_{\mathcal{H}}(\mu) \,\mathbb{1}_{2\times2}
+ \frac{\alpha_s}{4\pi} \left( \frac{1}{\eps^2} - \frac{1}{\eps} \ln \frac{- p_q^2 }{m_b u \mu} - \frac{1}{\eps} \ln \frac{-p_{\bar{q}}^2}{m_b \bar{u} \mu} - \frac{i \pi}{\eps}\right) \, \begin{pmatrix}
C_A & -1 \\
0 & 0
\end{pmatrix}
+ \mathcal{O}(\alpha_s^2) \,,
\end{equation}
where $\delta Z_{\mathcal{H}}(\mu)$ is given in~\eqref{eq:ZH}. Furthermore, we also kept the anti-hardcollinear quarks off-shell with $p_q^2 \equiv p_{q+} p_{q-} + i\eta$ and $p_{q+} = m_b u$ for the outgoing quark and $p_{\bar{q}}^2 \equiv p_{\bar{q}+} p_{\bar{q}-} + i\eta$ with $p_{\bar{q}+} = m_b \bar{u}$ for the outgoing antiquark, respectively. Soft-gluon exchanges between the two outgoing light-like directions generate a rescattering phase in~\eqref{eq:deltazsoft}.

Similar to soft exchanges, also hardcollinear momenta cannot change the $\mathcal{O}(1)$ momentum fractions of the anti-hardcollinear quarks. From the structure of the SCET operators in~\eqref{eq:SCETO12} it is also clear that the collinear contributions are identical to those from the operator $\mathcal{O}_{7\gamma}$,
\begin{align}
 \delta\mathbf{Z}_{\mathcal{O}_{12}}^{\rm hc}(u,v;\mu) = \delta(u-v) \, \delta Z_\chi(\mu) \, \mathbb{1}_{2\times 2} \,, 
\end{align}
with $\delta Z_\chi(\mu)$ given in~\eqref{eq:Zchi}.

\begin{figure}[t]
  \centering
  \includegraphics[width=0.7\textwidth]{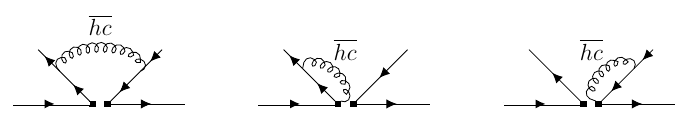}
  \caption{Anti-hardcollinear one-loop diagrams that contribute to the renormalisation of the operators $\mathcal{O}_{1,2}^q$.}
  \label{fig:O1_SCET_ren_anticollinear}
\end{figure}

Lastly, the singularity structure of anti-hardcollinear loops, shown in Fig.~\ref{fig:O1_SCET_ren_anticollinear}, is more intricate as the resulting expressions are non-local and need to be expanded in terms of plus-distributions in the momentum fraction $u$. Radiative corrections to the operator $\mathcal{O}_2^q$, for which the two anti-hardcollinear quarks form a colour singlet, are diagonal and result in the well-known Efremov-Radyushkin-Brodsky-Lepage evolution kernel~\cite{Lepage:1979zb,Lepage:1980fj,Efremov:1979qk} for the leading-twist light-cone distribution amplitude of a light pseudoscalar meson,
\begin{align}
\label{eq:VERBL}
&V_{\text{ERBL}}(u,v) \nonumber \\
= \, &\left[ \left(1+\frac{1}{v-u}\right) \frac{u}{v}\, \theta(v-u) + \left(1+\frac{1}{u-v}\right) \frac{1-u}{1-v} \,\theta(u-v) \right]_+^{(u)} \\[0.2em]
= \, &\left[ \left(1+\frac{1}{v-u}\right) \frac{u}{v}\, \theta(v-u) + \left(1+\frac{1}{u-v}\right) \frac{1-u}{1-v} \,\theta(u-v) \right]_+^{(v)} + \left( u \ln \bar{u} + \bar{u} \ln u + \frac32 \right) \delta(u-v) \,,  \nonumber 
\end{align}
where the superscript indicates whether the plus distribution is defined with respect to the variable $u$ or $v$, see discussion in~\cite{Beneke:2021pkl}. Corrections to the operator $\mathcal{O}_1^q$ induce a mixing with the operator $\mathcal{O}_2^q$. The resulting diagonal and off-diagonal contributions feature non-local poles $\sim V_{\rm ERBL}(u,v)$ as well as additional local contributions with logarithmic sensitivity to the off-shellness $q^2$ and $\bar{q}^2$ of the two anti-hardcollinear quarks. This is a consequence of the net colour charge in the anti-collinear sector and leads to a more complicated expression that is similar in structure to the QED evolution kernel of a charged pion~\cite{Beneke:2020vnb,Beneke:2021pkl},
\begin{align}
    \delta\mathbf{Z}_{\mathcal{O}_{12}}^{\rm \overline{hc}}(u,v;\mu) = - \, &\frac{\alpha_s}{4\pi} \left( \frac{2}{\eps^2} - \frac{1}{\eps} \ln \frac{-p_q^2}{\mu^2} - \frac{1}{\eps} \ln \frac{-p_{\bar{q}}^2}{\mu^2} + \frac{3}{2\eps} \right) \begin{pmatrix}
C_A & -1 \\
0 & 0
\end{pmatrix} \delta(u-v) \nonumber \\
- &\frac{\alpha_s}{4 \pi} \frac{1}{\eps} \begin{pmatrix}
2C_F - C_A & 1 \\
0 & 2 C_F
\end{pmatrix} V_{\text{ERBL}}(u,v) + \mathcal{O}(\alpha_s^2) \,.
\end{align}

Summing the soft, hardcollinear and anti-hardcollinear contributions, we find the renormalisation matrix for the effective operators $\mathcal{O}_{1,2}^q$,
\begingroup
\interdisplaylinepenalty=10000
\begin{align}
  &\mathbf{Z}_{\mathcal{O}_{12}}(u,v;\mu) \nonumber \\[0.2em]
  = \, &\mathbb{1}_{2\times 2} \, \delta(u-v) + \delta\mathbf{Z}_{\mathcal{O}_{12}}^{\rm soft}(u,v;\mu) + \delta\mathbf{Z}_{\mathcal{O}_{12}}^{\rm hc}(u,v;\mu) + \delta\mathbf{Z}_{\mathcal{O}_{12}}^{\rm \overline{hc}}(u,v;\mu) \, + \, \mathcal{O}(\alpha_s^2) 
  \nonumber \\[0.2em]
  = \, &Z_{\mathcal{O}_{7\gamma}}(\mu) \, \mathbb{1}_{2\times 2} \, \delta(u-v)
  - \frac{\alpha_s}{4 \pi} \frac{1}{\eps} \begin{pmatrix}
2C_F - C_A & 1 \\
0 & 2 C_F
\end{pmatrix} V_{\text{ERBL}}(u,v) \nonumber \\
- \, &\frac{\alpha_s}{4\pi} \left( \frac{1}{\eps^2} + \frac{1}{\eps} \ln \frac{\mu^2}{m_b^2 u \bar{u}} + \frac{i \pi}{\eps} + \frac{3}{2\eps} \right) \, \begin{pmatrix}
C_A & -1 \\
0 & 0
\end{pmatrix} \, \delta(u-v) + \mathcal{O}(\alpha_s^2) \,, \label{eq:Z12}
\end{align}
\endgroup
with $Z_{\mathcal{O}_{7\gamma}}(\mu)$ given in~\eqref{eq:Z7gamma}. Again, the dependence on the off-shellness $p^2$ of the collinear quark and $p_q^2$ as well as $p_{\bar{q}}^2$ for the anti-collinear quarks cancels, as expected.

\paragraph{Hard matching coefficients:}
\begin{figure}[t]
    \centering
     \includegraphics[width=0.75\textwidth]{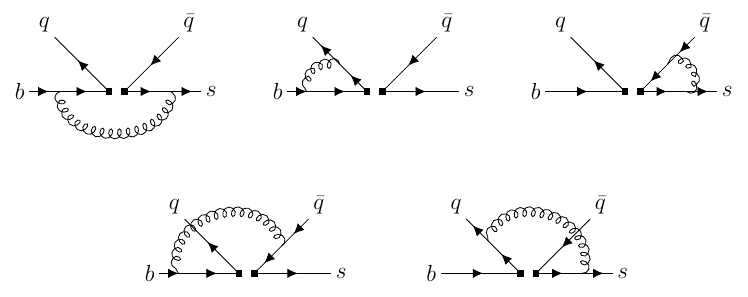}
    \caption{One-loop hard-matching corrections contributing to ${\bf H}_{12}(u;\mu,\nu)$.}
    \label{fig:hard_12}
\end{figure}
The hard-matching coefficients ${\bf H}_{12}(u;\mu,\nu)$ defined in~\eqref{eq:H12def} can be extracted from the $b \to s \bar{q} q$ amplitude with on-shell partons, where the outgoing quark (antiquark) carries fraction $u$ ($\bar{u} = 1-u$) of the photon momentum. At NLO, the diagrams in Fig.~\ref{fig:hard_12} together with the on-shell field renormalisation of the four external quarks yield the bare matrix elements with renormalised fields
\begin{align}
\label{eq:H1_bare_full}
    \begin{pmatrix}
        \langle Q_1^{q,\text{bare}}\rangle^{(1)} \\ \langle Q_2^{q,\text{bare}}\rangle^{(1)} \end{pmatrix}
        = \frac{\alpha_s}{4\pi} 
        \begin{pmatrix}
            C_F f_1(u) + C_A f_2(u) & -f_2(u) \\ -f_3(u) & C_F f_1(u) + C_A f_3(u)
        \end{pmatrix} 
        \begin{pmatrix}
        \langle \mathcal{O}_1^{q}\rangle^{(0)} \\ \langle \mathcal{O}_2^{q}\rangle^{(0)} \end{pmatrix} \,,
\end{align}
with the different loop functions defined as
\begin{equation}
f_1(u) =
-\frac{1}{\eps^2_{\IR}} - \frac{1}{\eps_{\IR}}\ln\frac{\mu^2}{m_b^2} - \frac{5}{2\eps_{\IR}}
- \frac{6}{\eps_{\UV}}
- \frac12 \ln^2 \frac{\mu^2}{m_b^2} - \frac{5}{2}\ln \frac{\mu^2}{m_b^2} - 6 \ln \frac{\nu^2}{m_b^2}
+ g(u) - \frac{\pi^2}{12} - 16 \,,
\end{equation}
and
\begin{align}
f_2(u) = &-\frac{1}{\eps_{\IR}^2}
- \frac{1}{\eps_{\IR}}
 \ln\!
\frac{\mu^2}{m_b^2 u\bar{u}}- \frac{i\pi}{\eps_{\IR}}
-\frac{3}{2\eps_{\IR}}+\frac{3}{\eps_{\UV}}
-\frac{1}{2} \ln^2\!
\frac{\mu^2}{m_b^2 u \bar{u}}
-\frac{3}{2} \ln \frac{\mu^2}{m_b^2}
+3 \ln \frac{\nu^2}{m_b^2} -i\pi\ln\frac{\mu^2}{m_b^2u\bar{u}} \nonumber \\[2pt]
&+\ln u\ln\bar u+\frac32 \ln^2\bar{u}+\frac{7\pi^2}{12}
+3-f(\bar{u})
-3 (1+i\pi)\ln\bar{u}\,,
\end{align}
and
\begin{equation}
f_3(u) = 
\frac{3}{\eps_{\UV}}
+3\ln\!\frac{\nu^2}{m_b^2}+5-\frac{1}{2}g(u) \,. 
\end{equation}
The two functions $f(u)$ and $g(u)$ are defined as~\cite{Beneke:2020vnb},
 \begin{align}
     f(u) = {\rm Li}_2(1-u) + 2 \ln^2 u - (3+2i\pi) \ln u -\frac{\ln u }{1-u}\,,
\label{eq:fdef}
 \end{align}
and
\begin{equation}
g(u) = 3 \left(\frac{1-2u}{1-u} \ln u - i \pi \right) + \bigg[2 \,{\rm Li}_2(u) - \ln^2 u + \frac{2 \ln{u}}{1-u} - (3+2 i\pi)\ln{u} - \,(u \to  1-u ) \bigg]\,,
\label{eq:gdef}
\end{equation}
and we furthermore set $E_\gamma = m_b/2$ in the above expressions.

As before, we separated UV singularities, which are cancelled by the renormalisation of $Q_{1,2}^q$ (see Appendix~\ref{app:evanescent_ops}) and the associated scale $\nu$ of the Wilson coefficients $C^q_{1,2}(\nu)$ from IR singularities and the factorisation scale $\mu$. The IR singularities match the UV poles of the SCET operators. 
As the LO hard matching functions are given by $\mathbf{H}_{12}(u;\mu,\nu) = \mathbb{1}_{2\times 2} + \mathcal{O}(\alpha_s)$, at NLO this amounts to subtracting the $\mathcal{O}(\alpha_s)$ term of
\begin{align}
  &\int_0^1 du'\, \mathbf{Z}_{\mathcal{O}_{12}}(u',u;\mu) \\[0.2em]
  = \, &\left[ 1
  - \frac{\alpha_s C_F}{4\pi} \left( \frac{1}{\eps^2} + \frac{1}{\eps} \ln \frac{\mu^2}{m_b^2} + \frac{5}{2\eps} \right) \right] \mathbb{1}_{2\times 2} 
- \frac{\alpha_s}{4\pi} \left( \frac{1}{\eps^2} + \frac{1}{\eps} \ln \frac{\mu^2}{m_b^2 u \bar{u}} + \frac{i \pi}{\eps} + \frac{3}{2\eps} \right) \, \begin{pmatrix}
C_A & -1 \\
0 & 0
\end{pmatrix} + \mathcal{O}(\alpha_s^2) \nonumber
\end{align}
from the (UV-renormalised) matrix in~\eqref{eq:H1_bare_full}. Note that here the contribution proportional to $V_{\rm ERBL}$ vanishes because the plus-distribution in the first line of~\eqref{eq:VERBL} acts on a constant.

Lastly, we quote the finite hard-matching functions, including the corrections arising from evanescent operators, which are discussed in greater detail in Appendix~\ref{app:evanescent_ops}:
\begin{align}
\label{eq:H12}
\mathbf{H}_{12}(u;\mu,\nu) =\;& \mathbb{1}_{2\times 2} + \frac{\alpha_s(\nu)}{4\pi}\Bigg\{
\ln\frac{\nu^2}{m_b^2}
\begin{pmatrix} 3/N_c & -3 \\ -3 & 3/N_c \end{pmatrix}
\nonumber \\
&+\left(-\frac{1}{2}\ln^2\frac{\mu^2}{m_b^2}-\frac{5}{2}\ln\frac{\mu^2}{m_b^2}
+g(u)-\frac{\pi^2}{12}-28\right) C_F\,\mathbb{1}_{2\times 2} \nonumber \\
&+\bigg(-\frac{1}{2}\ln^2\frac{\mu^2}{m_b^2u\bar u}-i\pi\ln\frac{\mu^2}{m_b^2u\bar u}-\frac{3}{2} \ln \frac{\mu^2}{m_b^2} +\ln u\ln\bar u + \frac32 \ln^2\bar{u}-3 (1+i\pi)\ln\bar{u} \nonumber \\
&\qquad +\frac{7\pi^2}{12}+7-f(\bar{u})\bigg)
\begin{pmatrix} C_A & -1 \\ 0 & 0 \end{pmatrix}
+\left(10-\frac{1}{2}g(u)\right)
\begin{pmatrix} 0 & 0 \\ -1 & C_A \end{pmatrix}
\Bigg\}\,.
\end{align}
For $\mu=\nu$, the linear combination $[\mathbf H_{12}]_{21}/N_c+[\mathbf H_{12}]_{22}$ reproduces the wrong-insertion relation of~\cite{Beneke:2009ek}. The one-loop matching matrix can also be reconstructed from the QED calculation of~\cite{Beneke:2020vnb} by replacing the electromagnetic charge factors with the corresponding QCD colour factors.

Including the Wilson coefficients of the weak effective Hamiltonian, the hard function entering the factorisation formula~\eqref{eq:factorisationtheorem} for the $Q_1^q$-$Q_{7\gamma}$ interference contribution is therefore
\begin{equation}
\label{eq:H1q}
   H_1^q(u;\mu) = C_1^q(\nu)\, \big[\mathbf{H}_{12}(u;\mu,\nu)\big]_{11}
   + C_2^q(\nu)\, \big[\mathbf{H}_{12}(u;\mu,\nu)\big]_{21} \,. 
\end{equation}

\paragraph{Anomalous dimensions:}
In analogy with~\eqref{eq:rge_H7}, the hard matching coefficients obey
\begin{equation}
    \frac{d}{d\ln\mu}\, \mathbf{H}_{12}(v;\mu,\nu) = \int_0^1 \! du\,  \mathbf{H}_{12}(u;\mu,\nu) \, \boldsymbol{\gamma}_{12}(u,v;\mu) \,,
    \qquad
    \boldsymbol{\gamma}_{12} \equiv - \frac{d\mathbf{Z}_{\mathcal{O}_{12}}}{d\ln\mu} \otimes \mathbf{Z}_{\mathcal{O}_{12}}^{-1} \,,
\end{equation}
with the one-loop anomalous dimension 
\begin{align}
\label{eq:Gamma12}
    \boldsymbol{\gamma}_{12}(u,v;\mu) =\;& \bigg\{ \gamma_J(\mu)\, \mathbb{1}_{2\times 2}
    - \bigg[ \frac{\gamma_{\rm cusp}(\alpha_s)}{2} \bigg( \ln\frac{\mu^2}{m_b^2 u\bar u} + i\pi \bigg) + 3\,\frac{\alpha_s}{4\pi} \bigg] \mathbf{T}_{\bar n}^2 \bigg\} \delta(u-v) \nonumber \\[2mm]
    &\, - \frac{\alpha_s}{\pi} \begin{pmatrix} C_F - C_A/2 & \,\,1/2 \\ 0 & \,\,C_F \end{pmatrix} V_{\rm ERBL}(u,v) + \mathcal{O}(\alpha_s^2) \,,
    \qquad
    \mathbf{T}_{\bar n}^2 = \begin{pmatrix} C_A & -1 \\ 0 & 0 \end{pmatrix} \,,
\end{align}
where $\gamma_J(\mu)$ is given in~\eqref{eq:Gamma7} and the constant term proportional to $\mathbf{T}_{\bar n}^2$ is the local one-loop non-cusp contribution from the anti-hardcollinear quark--antiquark sector. Further, in colour-space notation, $\mathbf{T}_{\bar n}^2$ represents the squared total colour charge of the anti-hardcollinear quark--antiquark pair in the $(\mathcal{O}_1^q, \mathcal{O}_2^q)$ basis with eigenvalue $0$ ($C_A$) for the colour-singlet (octet) configuration. The colour factor multiplying the cusp logarithm $\ln(\mu/m_b)$ is therefore the sum of the squared colour charges in the two collinear directions, $C_F\,\mathbb{1}+\mathbf{T}_{\bar n}^2$.

\section{Solutions to the renormalisation-group equations}
\label{sec:RGEsolution}
As the evolution of the weak effective operators with the renormalisation scale $\nu$ is well known, we focus in this section on solutions to the evolution equations in the factorisation scale $\mu$ only.

\subsection{Evolution of \texorpdfstring{$H_{7\gamma}$}{H7gamma}}
\label{subsec:evolution_H7}
The $\mu$ evolution of $H_{7\gamma}$ is governed by the standard anomalous dimension of a leading-power heavy-to-light SCET current. Evolution from the matching scale $\mu_0\sim m_b$ to $\mu$ is obtained by
\begin{equation}
\label{eq:H7_sol}
    H_{7\gamma}(\mu,\nu) = U_7(\mu_0,\mu)\, H_{7\gamma}(\mu_0,\nu) \,, 
    \qquad
    \ln U_7(\mu_0,\mu) = C_F \Big[ S(\mu_0,\mu) - a_\Gamma(\mu_0,\mu) \ln\frac{m_b}{\mu_0} \Big] - a_{\gamma'}(\mu_0,\mu)  \,.
\end{equation}
Up to a factor of two in the exponent, this expression coincides with the evolution factor $U_1$ in~\cite{Neubert:2004dd}, which governs the evolution of $|H_{7\gamma}|^2$ in the leading-power $Q_{7\gamma}$--$Q_{7\gamma}$ contribution. The functions $S(\mu_0,\mu)$ and $a_{\Gamma}(\mu_0,\mu)$ are defined by 
\begin{equation}
    S(\mu_0,\mu) = - \int_{\alpha_s(\mu_0)}^{\alpha_s(\mu)} \! d\alpha \, \frac{\gamma_{\rm cusp}(\alpha)}{\beta(\alpha)} \int_{\alpha_s(\mu_0)}^{\alpha} \frac{d\alpha'}{\beta(\alpha')} \,,
    \qquad
    a_{\Gamma}(\mu_0,\mu) = - \int_{\alpha_s(\mu_0)}^{\alpha_s(\mu)} \! d\alpha \, \frac{\gamma_{\rm cusp}(\alpha)}{\beta(\alpha)} \,,
\end{equation}
with $a_{\gamma'}$ obtained from $a_\Gamma$ by replacing
$\gamma_{\rm cusp}(\alpha)$ with $\gamma'(\alpha)$. Here, $\beta(\alpha_s) = d\alpha_s/d\ln\mu$ denotes the QCD $\beta$ function. At next-to-leading logarithmic (NLL) accuracy, the evolution functions take the form (see, e.g.,~\cite{Becher:2006mr})
\begin{align}
\label{eq:S_NLL}
    S(\mu_0,\mu) &= \frac{\gamma^{\rm cusp}_0 }{4\beta_0^2} \left[ \frac{4\pi}{\alpha_s(\mu_0)} \left( 1 - \frac{1}{r} - \ln r \right)
    + \left( \frac{\gamma^{\rm cusp}_1}{\gamma^{\rm cusp}_0} - \frac{\beta_1}{\beta_0} \right) (1 - r + \ln r) + \frac{\beta_1}{2\beta_0} \ln^2 r \right] \, \nonumber \\[0.2em]
    a_\Gamma(\mu_0,\mu) &= \frac{\gamma^{\rm cusp}_0 }{2\beta_0} \ln r  ,
    \qquad
    a_{\gamma'}(\mu_0,\mu) = \frac{\gamma'_0}{2\beta_0} \ln r \,,
\end{align}
where $r = \alpha_s(\mu)/\alpha_s(\mu_0)$ is evaluated using the two-loop running coupling. In addition to $\gamma^{\rm cusp}_0 = 4$ and $\gamma'_0 = -5C_F$, the required coefficients are $\gamma^{\rm cusp}_1 = \tfrac{4}{3}((4-\pi^2)C_A+5\beta_0)$, $\beta_0 = \tfrac{11}{3} C_A - \tfrac43 T_F n_f$ and $\beta_1 = \tfrac{34}{3} C_A^2 - \tfrac{20}{3} C_A T_F n_f - 4 C_F T_F n_f$.

\subsection{Evolution of \texorpdfstring{$H_{1}^q$}{H1^q}}
Since the anomalous dimension matrix in~\eqref{eq:Gamma12} is upper triangular, i.e. the $(2,1)$ entry of $(\boldsymbol{\gamma}_{12})$ vanishes, the hard function $H_1^q(u;\mu)$ in~\eqref{eq:H1q} that enters the factorisation formula for the $\bar{B} \to X_s \gamma$ decay rate obeys
\begin{equation}
\label{eq:RGE_H1q}
    \frac{d}{d\ln\mu}\, H_1^q(v;\mu) = \int_0^1 \! du\, \gamma_{1}(u,v;\mu) \, H_1^q(u;\mu) \,,
\end{equation}
with $\gamma_{1} \equiv (\boldsymbol{\gamma}_{12})_{11}$. Using $C_A = N_c$ and $2C_F - C_A = -1/N_c$, we write the corresponding anomalous dimension as
\begin{equation}
\label{eq:Gamma1}
    \gamma_{1}(u,v;\mu) = \widehat{\gamma}_1(\mu)\,\delta(u-v)
    + \frac{\alpha_s C_A}{2\pi} \left[\ln(u\bar u)\delta(u-v)\,  + \frac{1}{N_c^2}\, V_{\rm ERBL}(u,v) \right] + \mathcal{O}(\alpha_s^2) \,,
\end{equation}
with
\begin{align}
    \widehat{\gamma}_1(\mu) = -(C_F + C_A)\, \gamma_{\rm cusp}(\alpha_s) \ln\frac{\mu}{m_b} + \gamma'_1(\alpha_s) \,, \quad \gamma'_1(\alpha_s) = - \frac{\alpha_s}{4\pi} \big[ 5 C_F + (3 + 2i\pi)\, C_A \big] + \mathcal{O}(\alpha_s^2) \,.
\end{align}
The evolution kernel has a structure similar to that governing the QED-generalised light-cone distribution amplitude of a charged pion~\cite{Beneke:2020vnb,Beneke:2021pkl}. In both cases, the logarithmic dependence on the momentum fraction $u$ mixes Gegenbauer moments, so the evolution is no longer diagonal in Gegenbauer space.
In~\cite{Beneke:2021pkl}, these logarithmic terms are treated perturbatively in the electromagnetic coupling. Here, we instead treat the local logarithmic term exactly and expand in the colour-suppressed ERBL contribution, obtaining analytic expressions for the leading term and the first subleading correction in the large-$N_c$ expansion.

First, we write
\begin{equation}
    H_1^q(u;\mu) = \widehat{U}(\mu_0,\mu) \, h_1^q(u;\mu) \,,
\end{equation}
with
\begin{equation}
\label{eq:Uhat}
    \ln \widehat{U}(\mu_0,\mu) = (C_F + C_A)\Big[ S(\mu_0,\mu) - a_\Gamma(\mu_0,\mu) \ln\frac{m_b}{\mu_0} \Big] - a_{\gamma'_1}(\mu_0,\mu) \,,
\end{equation}
where $a_{\gamma'_1}$ is defined in analogy to $a_{\gamma'}$ in~\eqref{eq:S_NLL} and we retain the exact dependence on the colour factors in $\widehat{U}(\mu_0,\mu)$. With this redefinition, the function $h_1^q(u;\mu)$ evolves with the anomalous dimension 
\begin{equation}
    \gamma_{1}(u,v;\mu) - \widehat{\gamma}_1(\mu)\,\delta(u-v)
    = \frac{\alpha_s C_A}{2\pi} \left[\ln(u\bar u)\delta(u-v)\,  + \frac{1}{N_c^2}\, V_{\rm ERBL}(u,v) \right] + \mathcal{O}(\alpha_s^2) \,.
\end{equation}
We solve the resulting evolution equation perturbatively in the colour-suppressed ERBL term and define the leading-colour (LC) and next-to-leading-colour (NLC) coefficients by
\begin{equation}
    h_1^q(u;\mu)
    = h_1^q(u;\mu) \Big\vert_{\rm LC}
    + \frac{1}{N_c^2}\,h_1^q(u;\mu) \Big\vert_{\rm NLC}
    + \mathcal{O}(N_c^{-4}) \,.
\end{equation}
The solution for the leading term reads
\begin{align}
   h_1^q(u;\mu)\Big\vert_{\rm LC} = (u\bar u)^{2C_A a}\, h_1^q(u;\mu_0)\Big\vert_{\rm LC} \,, \qquad \text{with} \qquad a \equiv a(\mu_0,\mu) = -\frac{1}{2\beta_0} \ln r \,.
\end{align}
The next term satisfies
\begin{align}
   \frac{d}{d\ln\mu}\,h_1^q(u;\mu)\Big\vert_{\rm NLC}
   ={}& \frac{\alpha_s C_A}{2\pi}\ln(u\bar u)\,
   h_1^q(u;\mu)\Big\vert_{\rm NLC}
   \nonumber\\
   &+\frac{\alpha_s C_A}{2\pi}\int_0^1\!dv\,
   V_{\rm ERBL}(v,u)\,
   (v\bar{v})^{2C_A a}\,
   h_1^q(v;\mu_0)\Big\vert_{\rm LC} \,,
\end{align}
and the solution can be written as
\begin{align}
   h_1^q(u;\mu)\Big\vert_{\rm NLC}
   = (u\bar u)^{2C_A a}\,
   h_1^q(u;\mu_0)\Big\vert_{\rm NLC}
   +\int_0^1\!dv\,V_{\rm ERBL}(v,u)\,
   \frac{(v\bar v)^{2C_A a}-(u\bar u)^{2C_A a}}
        {\ln(v\bar v)-\ln(u\bar u)}\,
   h_1^q(v;\mu_0)\Big\vert_{\rm LC} \,.
\end{align}
We thus obtain the following approximate analytic solution to the RG equation of the hard function,
\begin{align}
\label{eq:LCandNLC}
    H_1^q(u;\mu)
    = & \widehat{U}(\mu_0,\mu)\,(u\bar u)^{2C_A a}
    \Bigg\{
    H_1^q(u;\mu_0)
    \nonumber\\
    &\qquad\qquad
    + \frac{1}{N_c^2}\int_0^1\!dv\,V_{\rm ERBL}(v,u)\,
    \frac{\bigl[v\bar v/(u\bar u)\bigr]^{2C_A a}-1}
         {\ln(v\bar v)-\ln(u\bar u)}\,
    H_1^q(v;\mu_0)
    \Bigg\}
    + \mathcal{O}(N_c^{-4}) \,.
\end{align}

\begin{figure}[t]
    \centering
    \includegraphics[width=\textwidth]{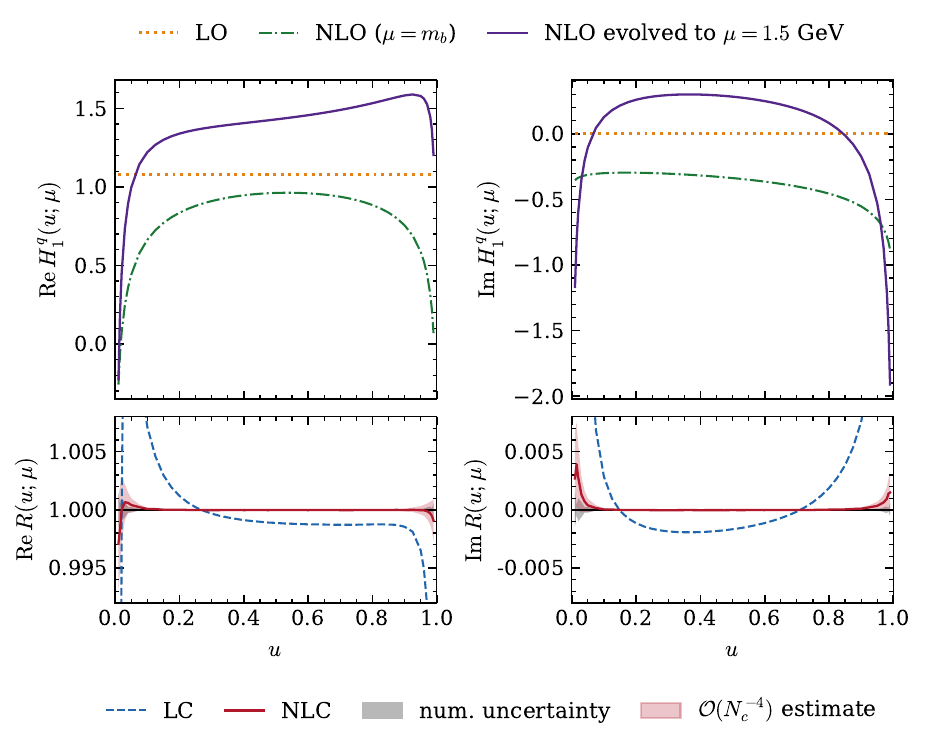}
    \vspace{-0.85cm}
    \caption{Real (left) and imaginary (right) parts of the hard function $H_1^q(u;\mu)$. The upper panels show the LO result (dotted), the NLO result at $\mu = \nu = m_b$ (dash-dotted), and the NLO result evolved to $\mu = 1.5$~GeV at NLL accuracy, obtained from the numerical solution of~\eqref{eq:RGE_H1q} in Gegenbauer space (solid). The lower panels show the ratio $R(u;\mu)$ of the LC approximation (first term in~\eqref{eq:LCandNLC}, dashed) or the NLC approximation (full expression in~\eqref{eq:LCandNLC}, solid) to the solution of the full-colour RG equation at $\mu = 1.5$~GeV. The grey band indicates the uncertainty of the full-colour solution from the truncation of the Gegenbauer expansion, the red band the uncertainty of the NLC result, estimated by the size of the $\mathcal{O}(N_c^{-4})$ term. We use $m_b = 4.8$~GeV, $N_c = 3$, and $\alpha_s(m_Z) = 0.118$ with two-loop running ($n_f = 5$ above and $n_f = 4$ below $m_b$). For the Wilson coefficients, we use the NLO values of~\cite{Beneke:2001ev}, evolved to $\nu = m_b$, $C_1^q = 1.073$ and $C_2^q = -0.173$.}
    \label{fig:H1q_evolution}
\end{figure}

To assess the accuracy of these approximations, Fig.~\ref{fig:H1q_evolution} compares the LC and NLC results with a numerical solution of the full-colour RG equation~\eqref{eq:RGE_H1q} at $\mu = 1.5$~GeV. We obtain the numerical solution by expanding $H_1^q$ in Gegenbauer polynomials $C_n^{3/2}(2u-1)$. The logarithmic dependence of the evolution kernel on $u$ induces mixing among the Gegenbauer coefficients, described by an infinite-dimensional matrix, as in the analysis of~\cite{Beneke:2021pkl}. For the results shown in Fig.~\ref{fig:H1q_evolution}, we retain the first 450 Gegenbauer polynomials and estimate the truncation uncertainty from the difference relative to the result obtained with 300 polynomials. In the LC and NLC approximations, we retain the full colour dependence of the initial condition at $\mu_0$, the evolution factor $\widehat U$ in~\eqref{eq:Uhat}, and the running coupling, and expand only the remaining $u$-dependent evolution according to~\eqref{eq:LCandNLC}. Lastly, the uncertainty of the NLC approximation is estimated by the size of the next, $\mathcal{O}(N_c^{-4})$, term in the expansion, which we evaluate numerically.

The LC approximation deviates from the numerical solution by $\sim 2$\textperthousand{} in the central region and by up to $12\%$ near the endpoints. Including the NLC term brings the analytic and numerical solutions into sufficient agreement for practical purposes. Their relative difference is approximately $3 \times 10^{-5}$ in the central region and reaches $4 \times 10^{-3}$ at $u = 0.01$. This good convergence reflects the combination of colour suppression and the smallness of the evolution variable. The ERBL correction in~\eqref{eq:LCandNLC} carries an overall suppression factor $2|a|/N_c \simeq 0.019$, with $|a| \simeq 0.028$ for $\mu_0 = m_b$ and $\mu = 1.5$~GeV.

\section{Radiative charm-loop jet function at NLO for \texorpdfstring{$m_c \to 0$}{mc → 0}}
\label{sec:antijet_function}
The anti-hardcollinear jet function $\bar{J}(u,\omega_1;\mu)$ describes the conversion of the quark-antiquark pair emitted from the four-quark operator $\mathcal{O}_{1}^q$ into an energetic photon and a soft, transversely polarised gluon, see Fig.~\ref{fig:antijet_LO_diagrams}. It enters the factorisation formula~\eqref{eq:factorisationtheorem} through a convolution with the hard function $H_1^q(u;\mu)$ in the momentum fraction $u$ of the emitted quark relative to the photon, and with the subleading shape function $g_{17}(\omega,\omega_1;\mu)$ in the soft light-cone momentum $\omega_1$ of the gluon. In this section, we compute this subleading-power radiative jet function at NLO and in the massless limit, $m_c \to 0$, which is smooth. Including the full charm-mass dependence is necessary for phenomenology. However, it introduces an additional scale to the integrals and makes the calculation substantially more challenging. As we aim to lay out the theoretical framework in the present paper, we leave this calculation for future work.

\subsection{Definition}
\label{sec:antijet_extraction}
In the $\bar{B} \to X_s \gamma$ decay rate, the diagrammatic cuts pass through the energetic strange-quark propagator because the strange quark forms part of the hadronic final state $X_s$. The anti-hardcollinear charm loop remains uncut, and the corresponding radiative jet function is therefore defined at the amplitude level and can be extracted from the partonic $b \to s \gamma g$ decay induced by the four-quark operators $Q_{1,2}^q$. Starting from the matching relation~\eqref{eq:H12def}, we write at LO in the electromagnetic coupling 
\begin{align}
  \big\langle s \gamma g \big| Q_{i}^q(0) \big|b\big\rangle &= \int_0^1 \! du \, [\mathbf{H}_{12}]_{i1} \, \big\langle s \gamma g \big| T\Big\{\tilde{\mathcal O}_1^q(u), i\!\int\! d^4y\, \mathcal{L}^{(1)}_{\xi}(y) \Big\} \big|b_v \big\rangle \,, 
\end{align}
where the effective SCET operator $\tilde{\mathcal{O}}_1^q$ appears in a time-ordered product with the subleading Lagrangian~\cite{Beneke:2002ni}
\begin{align}
\label{eq:Lxi1}
  \mathcal{L}_\xi^{(1)}(y)=\bar{\chi}_{\overline{hc}}(y)\, i y^\mu_\perp \big[i\bar n\cdot\partial\, \mathcal{A}_{\perp\mu,s}\big](y_+)\, \frac{\nslash}{2}\, {\chi}_{\overline{hc}}(y) \,,
\end{align}
where $\mathcal{A}^\mu_{s} \equiv S_{\bar n}^\dagger\,[iD_s^\mu S_{\bar n}]$ is the gauge-invariant soft gluon building block. It describes the interaction of a transversely polarised soft gluon with an energetic quark, an interaction absent at leading power. Note that the contribution from $\tilde{\mathcal{O}}_2^q$ vanishes to all orders due to the colour-singlet projection of the two anti-hardcollinear quark fields, see~\eqref{eq:SCETO12}.

In the next step, we integrate out the anti-hardcollinear fluctuations and match the time-ordered product in SCET-1 to the low-energy theory SCET-2 which contains only soft, collinear and anticollinear fields describing the external on-shell particles in the amplitude.
Defining subleading-power radiative jet functions through such a matching relation avoids subtleties related to their gauge invariance~\cite{Boer:2023yde}, which were pointed out in~\cite{Bodwin:2023asf}. Based on boost invariance, conservation of mass dimension and power counting of the fields, it is straightforward to show that only two operator structures are allowed in SCET-2,
\begin{align}
\label{eq:Tproductmatching}
  &T\Big\{\tilde{\mathcal O}_1^q(u), i\!\int\! d^4y\, \mathcal{L}^{(1)}_{\xi}(y) \Big\}
  = \int_{-\infty}^\infty \! d\omega_1 \left(\bar{j}(1-u,\omega_1) \tilde{\mathcal{O}}^a(\omega_1) -\bar{j}(u,\omega_1) \tilde{\mathcal{O}}^b(\omega_1) \right) \,. 
\end{align}
These two operators only differ by the respective order of Dirac matrices contracted with the photon field $A_{\perp,\overline{c}}^{\rm em}$ and the gluon field, represented by its gauge-invariant building block $\mathcal{A}_{\perp,s}$,
\begin{align}
\label{eq:OaOb}
    \tilde{\mathcal{O}}^a(\omega_1) &= Q_q \int \frac{dr}{2\pi}\, e^{-i \omega_1 r} \, \bar{\chi}^{(s)}_{c}\, \frac{\slashed{\bar n}}{2}\, \slashed{\cal A}_{\perp,s}(r\bar{n})\, [i n \cdot \partial \slashed{A}_{\perp,\overline{c}}^{\rm em}]\, (1-\gamma_5)\, h_v \,, \nonumber \\
    \tilde{\mathcal{O}}^b(\omega_1) &= \,Q_q \int \frac{dr}{2\pi}\, e^{-i \omega_1 r} \, \bar{\chi}^{(s)}_{c}\, \frac{\slashed{\bar n}}{2}\, [i n \cdot \partial \slashed{A}_{\perp,\overline{c}}^{\rm em}] \, \slashed{\cal A}_{\perp,s}(r\bar{n})\, (1-\gamma_5)\, h_v \,,
\end{align}
and charge conjugation symmetry imposes a relation between the two matching functions, already implemented in~\eqref{eq:Tproductmatching}. After integration with the LO hard function $[\mathbf{H}_{12}]_{11}^{(0)} = 1$,
\begin{align}
  &\int_0^1 \! du \, T\Big\{\tilde{\mathcal O}_1^q(u), i\!\int\! d^4y\, \mathcal{L}^{(1)}_{\xi}(y) \Big\}
  = \int_{-\infty}^\infty \! d\omega_1 \int_0^1 \! du \, \bar{j}(u,\omega_1) \left(\tilde{\mathcal{O}}^a(\omega_1) - \tilde{\mathcal{O}}^b(\omega_1) \right) \,, 
\end{align}
this reduces to one operator containing the commutator of the gluon and photon field. 

Finally, multiplying with the complex-conjugate amplitude of the weak operator $Q_{7\gamma}$ and performing the polarisation sum for the on-shell photon leads to the  jet function $\bar{J}(u,\omega_1)$ entering the factorisation formula for the $\bar{B} \to X_s \gamma$ decay rate~\eqref{eq:factorisationtheorem}. The relevant linear combination reads
\begin{equation}
\label{eq:antijet_def}
  \bar{J}(u,\omega_1;\mu) = \frac{2E_\gamma}{\omega_1+i\eta}\,
  \Big[2(1-\epsilon)\,\bar{j}(u,\omega_1) - 2\epsilon\,\bar{j}(1-u,\omega_1)\Big]\,,
\end{equation}
where $2E_\gamma = n\cdot k_\gamma$ arises from  $[in\cdot\partial \slashed{A}^{\rm em}_{\perp,\overline{c}}]$ in~\eqref{eq:OaOb}, and the factor $1/\omega_1$ arises from expressing the gauge-invariant soft gluon field $\mathcal{A}_{\perp,s}$ in~\eqref{eq:OaOb} in terms of the gluon field-strength tensor, such that $\bar{J}$ agrees with the normalisation of~\cite{Bartocci:2026eeq} and matches the definition of the subleading shape function $g_{17}$ in~\eqref{eq:Q17}.

\subsection{Strategy of the calculation and LO result}
\label{sec:strategyandLO}
\begin{figure}[t]
    \centering
    \includegraphics[width=0.75\textwidth]{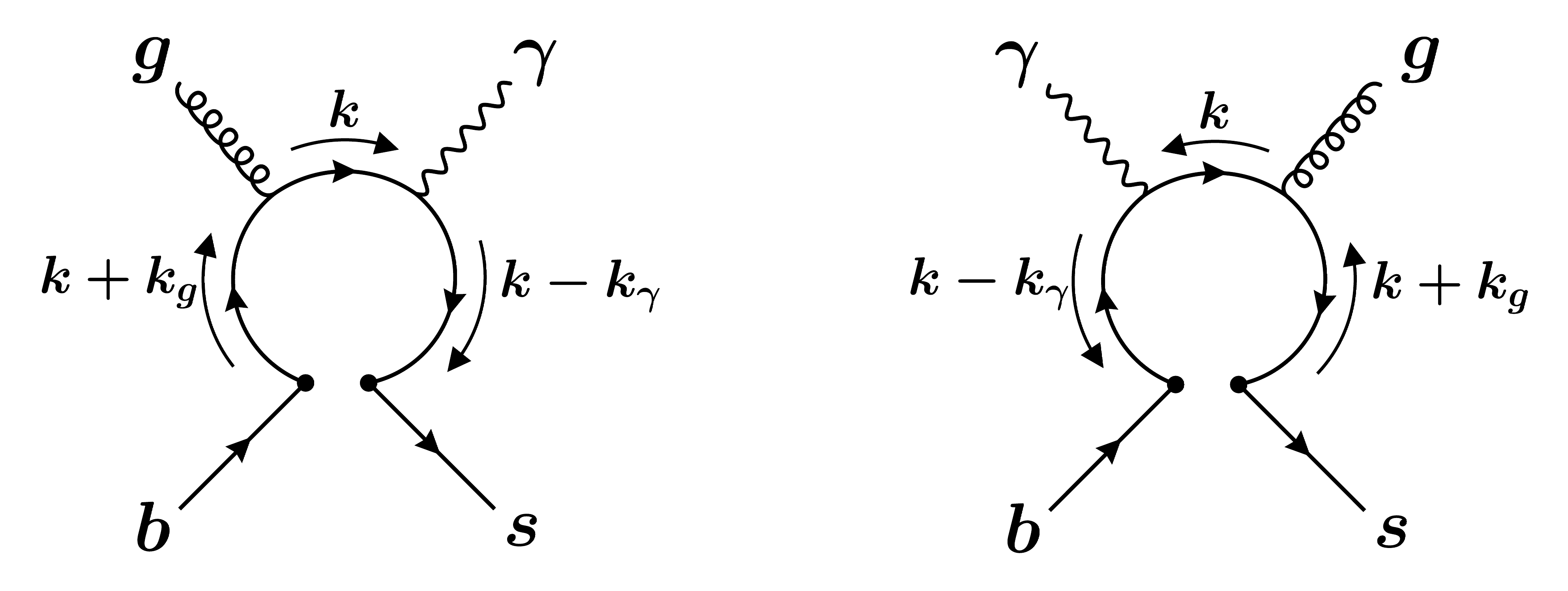}
    \caption{LO diagrams contributing to the $b \to s\gamma g$ amplitude. The displacement at the four-quark interaction vertex represents the non-locality of the anti-hardcollinear quark fields along the light-cone, reflected by the Feynman rule in~\eqref{eq:nonlocal_Feynmanrule}.}
    \label{fig:antijet_LO_diagrams}
\end{figure}

In practice, instead of using SCET Feynman rules for the calculation of the time-ordered product, it is simpler to extract $\bar{J}(u,\omega_1)$ from a method-of-regions~\cite{Beneke:1997zp} analysis of QCD diagrams. To this end, we consider on-shell external particles with momenta $b(p_b) \to s(p_s) \gamma(k_\gamma) g(k_g)$, isolate the purely anti-hardcollinear momentum region of the corresponding one- and two-loop diagrams shown in Figs.~\ref{fig:antijet_LO_diagrams},~\ref{fig:antijet_NLO_diagrams_1} and~\ref{fig:antijet_NLO_diagrams_2}, and identify the matrix elements of the two SCET-2 operators in~\eqref{eq:Tproductmatching}. In the following, we compute $\mathcal{O}(\alpha_s)$ corrections to $\bar{J}(u,\omega_1)$ using this method-of-region approach, but we have explicitly checked that the result agrees with a SCET calculation at the one-loop level.

In the limit $m_c \to 0$, the only non-vanishing dimensionful Lorentz invariant characterising the anti-hardcollinear momentum region is the scalar product of the soft-gluon momentum $k_g$ and the photon momentum $k_\gamma$, with $k_\gamma \cdot k_g = E_\gamma \omega_1$. The dependence on the soft momentum variable $\omega_1$ is therefore fixed by dimensional analysis in each graph in dimensional regularisation. Furthermore, the directed flow of positive large light-cone momentum implies that $\bar{J}(u,\omega_1)$ is analytic in the upper half of the complex $\omega_1$ plane, see also the discussion in~\cite{Bartocci:2024bbf}. The prescription $\omega_1 \to \omega_1 + i\eta$ is therefore understood throughout the following.

The dependence on the momentum fraction $u$ arises from the non-locality of the anti-hardcollinear quark fields in the SCET operator $\mathcal{O}_1^q(s)$ and is incorporated in the calculation through a corresponding Feynman rule for the four-quark vertex,
\begin{equation}
\label{eq:nonlocal_Feynmanrule}
  \raisebox{-0.5\height}{\includegraphics[width=3cm]{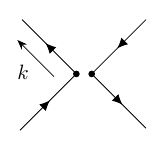}}
  \;=\;
  \gamma_\mu (1-\gamma_5)\otimes\gamma^\mu (1-\gamma_5)\;
  \delta\!\left(u-\frac{n\cdot k}{2E_\gamma}\right).
\end{equation}
The delta function fixes the fraction $u \in [0,1]$ of the photon's large light-cone momentum, $n\cdot k_\gamma = 2E_\gamma$, carried by the quark. The antiquark therefore carries the fraction $\bar u\equiv 1-u$.

\paragraph{LO result:} 
The two one-loop diagrams contributing at leading order are shown in Fig.~\ref{fig:antijet_LO_diagrams} and were already computed in~\cite{Bartocci:2026eeq}. For on-shell particles, the scalar product $2 k_\gamma \cdot k_g = 2 E_\gamma \omega_1$ is the only dimensionful scale present in these diagrams, and hence their anti-hardcollinear region coincides with the full amplitude. The sum of the two diagrams yields
\begin{align}
\label{eq:one_loop_massless_udep}
    \bar{j}^{(0)}(u,\omega_1) = \frac{1}{(4\pi)^2}
  \left(\frac{-2E_\gamma \omega_1}{\mu^2}\right)^{-\eps}
 \times f_{\text{LO}}(u)\,,
\end{align}
with
\begin{equation}
  f_{\text{LO}}(u) = 2 e^{\gamma_E\eps} \frac{\Gamma(1+\eps)}{1-\eps} (u \bar{u})^{-\eps}\left[u+\eps\bar{u}\right] = 2u + \mathcal{O}(\eps) \,.
\end{equation}
Using~\eqref{eq:antijet_def}, we then obtain the bare LO result for the radiative charm-loop jet function,
\begin{align}
\label{eq:antijet_LO_udep}
  \bar{J}^{(0)}(u,\omega_1) = \frac{1}{2\pi^{2}} \frac{E_\gamma}{\omega_1}
  \left(\frac{-2E_\gamma \omega_1 u \bar{u}}{\mu^2}\right)^{-\eps}
  e^{\gamma_E\eps}\,\frac{\Gamma(1+\eps)}{1-\eps}\,
\big(u (1-\eps) - \eps^2\big) 
= \frac{E_\gamma u}{2\pi^{2} \omega_1} + \mathcal{O}(\eps)\,,
\end{align}
such that
\begin{align}
    \int_0^1 \! du \, \bar{J}^{(0)}(u,\omega_1) = \frac{1}{2\pi^{2}} \frac{E_\gamma}{\omega_1}
  \left(\frac{-2E_\gamma \omega_1}{\mu^2}\right)^{-\eps}
  e^{\gamma_E\eps}\,(1-\eps-2\eps^2) \frac{\Gamma(1-\eps)^2\Gamma(1+\eps)}{\Gamma(3-2\eps)}
 = \frac{E_\gamma}{4 \pi^2 \omega_1} + \mathcal{O}(\eps) \,,
\end{align}
after integration with the LO hard function, which is constant in $u$.

\subsection{NLO calculation}
\label{sec:antijet_NLO_structure}
\begin{figure}[t]
    \centering
    \includegraphics[width=0.75\textwidth]{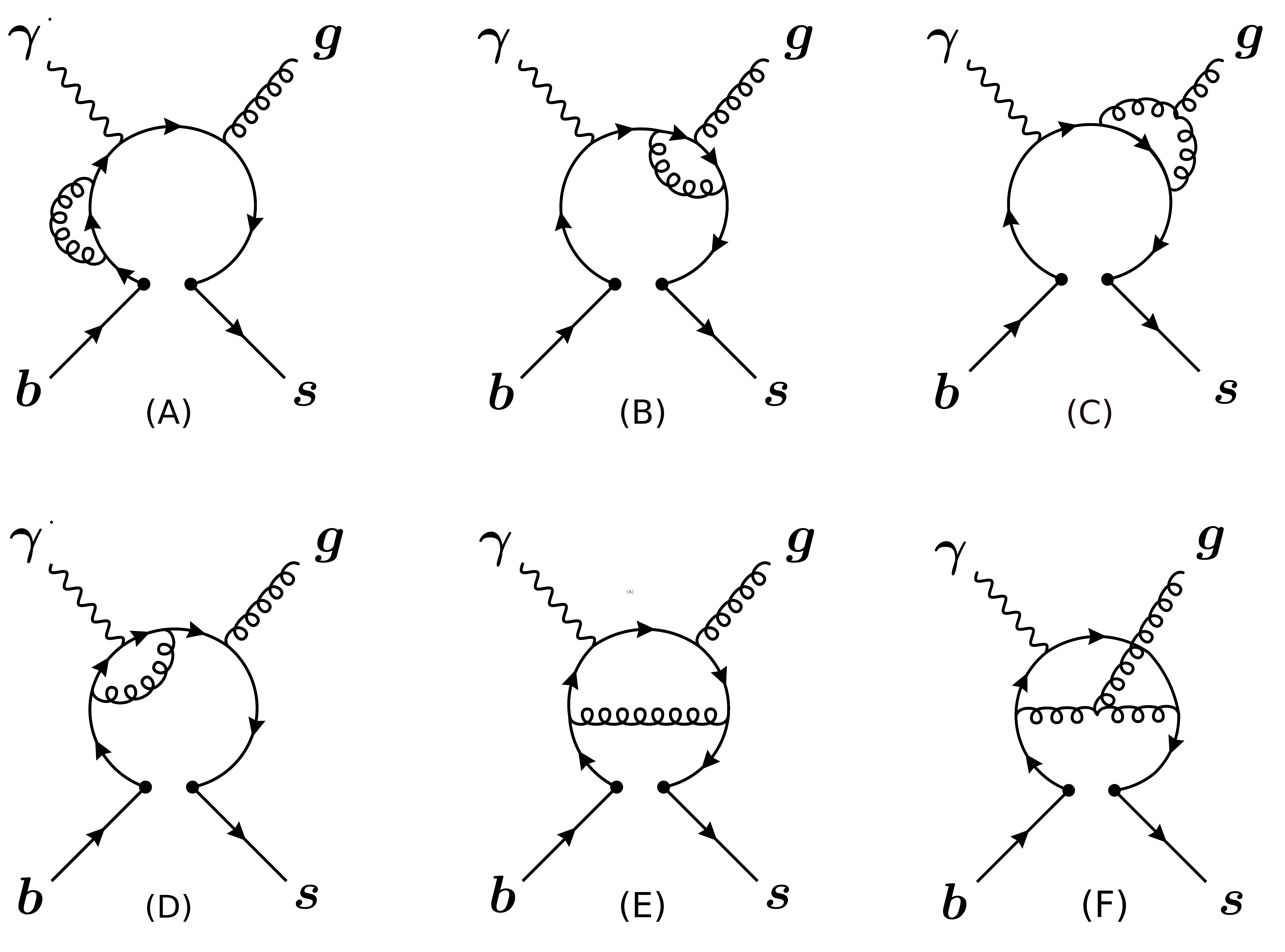}
\caption{Sample two-loop diagrams contributing to the $b \to s\gamma g$ amplitude at NLO. The diagrams shown here do not involve a gluon connection to the external bottom or strange quark line. Each of the diagrams labelled (A)--(F) represents a set of related diagrams. In (A), there are three possible insertions of the quark self-energy. Including the two orderings of the external photon and gluon gives six diagrams. For (B)--(E), the opposite ordering of the photon and gluon emissions gives one additional diagram in each class. In (F), only the diagram shown contributes. The total number of diagrams is therefore $6+2+2+2+2+1=15$.}
    \label{fig:antijet_NLO_diagrams_1}
\end{figure}

\begin{figure}[t]
    \centering
    \includegraphics[width=0.75\textwidth]{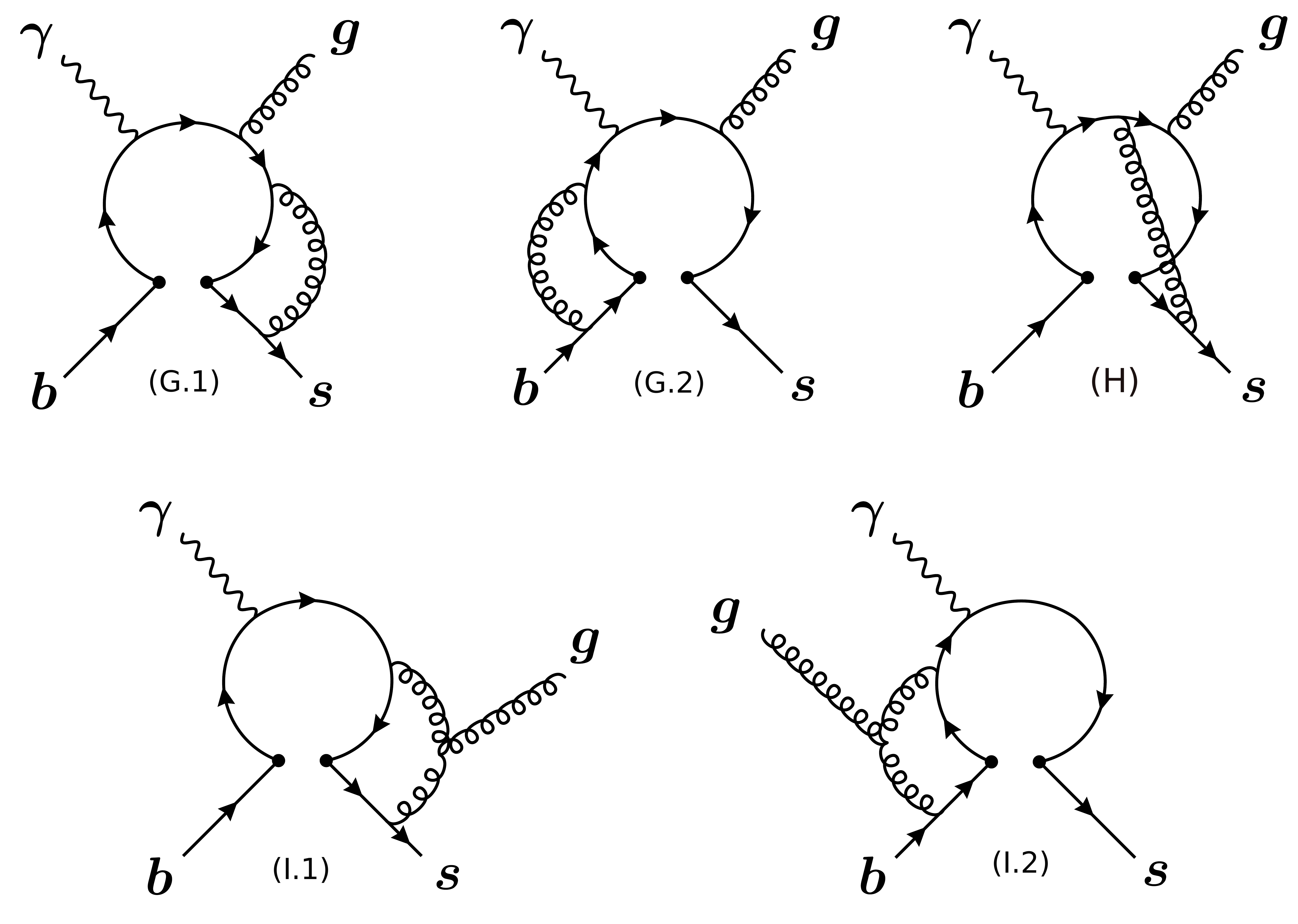}
    \caption{The NLO diagrams shown here do involve a gluon that connects to the external bottom or strange quark line with an eikonal coupling. For each of the diagrams labelled (G.1)--(I.2), a second diagram is obtained by attaching the gluon to the other external quark line. For (G.1), (G.2), and (H), interchanging the external photon and gluon doubles the number of diagrams in each set from two to four. Together with the two diagrams associated with each of (I.1) and (I.2), this gives $4+4+4+2+2=16$ diagrams in total.}
    \label{fig:antijet_NLO_diagrams_2}
\end{figure}

At NLO, we distinguish two classes of diagrams, with representatives shown in Figs.~\ref{fig:antijet_NLO_diagrams_1}
and~\ref{fig:antijet_NLO_diagrams_2}. Diagrams (A)--(F), displayed in Fig.~\ref{fig:antijet_NLO_diagrams_1}, involve gluon exchanges entirely within the quark loop, with no gluon attachments to the external bottom or strange quark lines. As at LO, only the anti-hardcollinear momentum region contributes to these diagrams. Diagrams (G)--(I), displayed in Fig.~\ref{fig:antijet_NLO_diagrams_2}, instead involve gluon attachments to the external bottom or strange quark lines. Here, expanding in the region where both loop momenta are anti-hardcollinear renders the couplings to the external quark lines eikonal. Let $l$ denote the gluon momentum flowing into the respective quark line. We can then approximate
\begin{equation}
    \bar{u}_s(p_s) \, (i g_s \slashed{A}_{\overline{hc}}) \, \frac{i(\slashed{p}_s - \slashed{l})}{(p_s - l)^2} \dots \, \simeq \, \bar{u}_\xi \, \frac{g_s n \cdot A_{\overline{hc}}}{n \cdot l} \dots \,,  
\end{equation}
for the strange-quark line, with $\bar{u}_\xi$ the leading component of the on-shell strange-quark spinor in the high-energy limit, and 
\begin{equation}
    \dots \frac{i(\slashed{p}_b + \slashed{l} + m_b)}{(p_b + l)^2 - m_b^2} \, (i g_s \slashed{A}_{\overline{hc}}) \, u(p_b) \, \simeq \, \dots \frac{-g_s n \cdot A_{\overline{hc}}}{n \cdot l} \, u_v \,,  
\end{equation}
for the heavy-quark line, with $u_v$ the leading component of the $b$-quark spinor in the heavy-quark limit. This reflects that attachments to the external heavy and collinear lines are described by anti-hardcollinear Wilson lines
\begin{equation}
W_{\overline{hc}}(x) = P\exp\left[ig_s\int_{-\infty}^0 ds\,
n\cdot A_{\overline{hc}}(x+sn)\right] \,,
\end{equation}
which are part of the gauge-invariant building blocks for the anti-hardcollinear quark fields in the SCET operator $\mathcal{O}_1^q$. The corresponding eikonal couplings vanish in light-cone gauge, $n \cdot A_{\overline{hc}} = 0$. In the following, however, we perform the calculation in Feynman gauge. As a consequence of the eikonal couplings, the terms proportional to $C_F$ cancel upon summing the two possible gluon attachments to the external quark lines in Fig.~\ref{fig:antijet_NLO_diagrams_2}, leaving only terms proportional to $C_A$.

As discussed above, the dependence on the dimensionful soft light-cone momentum $\omega_1$ is determined from dimensional analysis. We hence define at NLO
\begin{align}
\label{eq:two_loop_massless_udep}
  \bar{j}^{(1)}(u,\omega_1) =   \frac{1}{(4\pi)^2}\,\frac{\alpha_s}{4\pi}
  \left(\frac{-2E_\gamma \omega_1}{\mu^2}\right)^{-2\eps} f_{\text{NLO}}(u) \,.
\end{align}
In the following, we compute the function $f_{\text{NLO}}(u)$ from the two-loop diagrams shown in Figs.~\ref{fig:antijet_NLO_diagrams_1} and~\ref{fig:antijet_NLO_diagrams_2}. In Ref.~\cite{Bartocci:2026eeq}, some of us presented results for the integrated quantity $\int_0^1 du f_{\text{NLO}}(u)$. Here, we retain the full dependence on the momentum fraction $u$. This introduces an additional dimensionless parameter into the calculation, thereby increasing both the number of master integrals and their complexity.

\begin{figure}[t]
    \centering
    \includegraphics[width=1\textwidth]{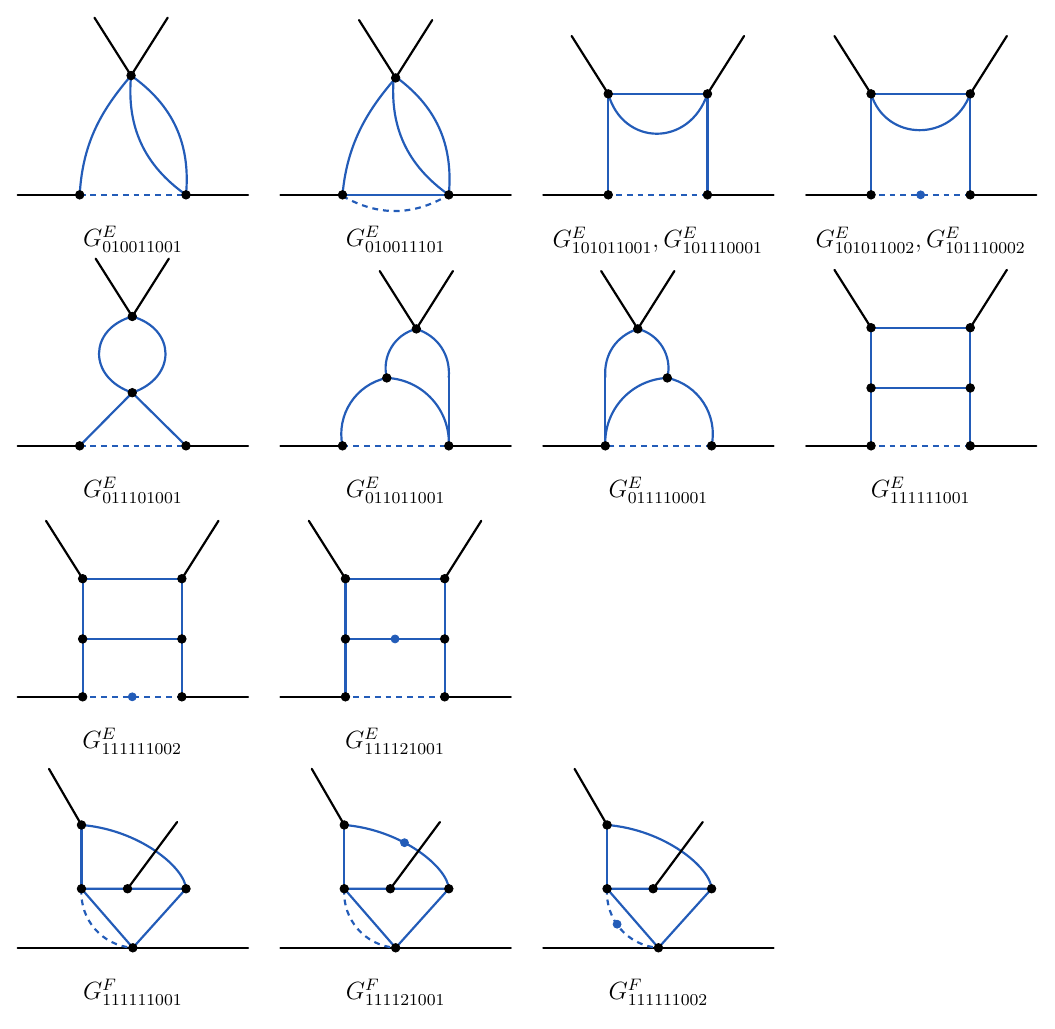}
    \caption{Master integrals for both topologies. The blue solid lines represent massless propagators while the black lines correspond to external particles. The dashed line represents the linear propagator from the energy constraint after using reverse unitarity, and a blue dot on a line indicates a squared propagator.} 
    \label{fig:MI_diagrams}
\end{figure}

We apply the same reduction procedure to all diagrams. First, tensor reduction is performed with
\texttt{FeynCalc}~\cite{Shtabovenko:2016sxi,Shtabovenko:2020gxv,Shtabovenko:2023idz,Mertig:1990an} followed by IBP reduction with \texttt{FIRE6}~\cite{Smirnov:2019qkx}, supplemented by
\texttt{LiteRed}~\cite{Lee:2012cn} to detect additional symmetries among the integrals. In this way each amplitude is expressed as a linear combination of
a set of master integrals. 
In order to treat the delta function from the Feynman rule in~\eqref{eq:nonlocal_Feynmanrule} we use reverse unitarity~\cite{Anastasiou:2002yz} to convert it to a difference of linear propagators, 
\begin{equation}
    \label{eq:ReverseUnitarity}
    -2\pi i \,\delta\left(n\cdot k-2E_\gamma u
    \right)=\frac{1}{n\cdot k-2E_{\gamma}u+i \eta}-\frac{1}{n\cdot k-2E_{\gamma}u-i \eta}\,,
\end{equation}
and make use of the fact that the integral reduction is inert to the sign of the $i \eta$ prescription.
We define the master integrals belonging to a generic topology $T$ as
\[
G^{T}_{a_1\ldots a_n} =
\int \frac{d^D l\, d^D k}{(2\pi)^{2D}}
\prod_{i=1}^{n} D_i^{a_i}\,,
\]
where $D_i$ are the elements of a topology. We find that all master integrals fall into either of two topologies $T=E,F$, defined as
\begin{align}
\label{eq:topD}
    E = \Bigg\{& \frac{1}{l^2 + i\eta},\,\frac{1}{(l+k_g)^2 + i\eta},\, \frac{1}{(k + k_g)^2 + i\eta},\,\frac{1}{(l - k_\gamma)^2 + i\eta}, \,\frac{1}{(k - l)^2 + i\eta},\frac{1}{(k - k_\gamma)^2 + i\eta}, \nonumber \\
    &\,
    \,\frac{1}{k^2 + i\eta}, \,\frac{1}{n\cdot l + i\eta},\, \frac{1}{n\cdot k-2E_{\gamma}u+ i\eta}\Bigg\} \,,
\end{align}
and
\begin{align}
\label{eq:topE}
    F = \Bigg\{& \frac{1}{l^2 + i\eta},\,\frac{1}{k^2 + i\eta},\, \frac{1}{(l - k_g)^2 + i\eta},\,\frac{1}{(k-l)^2 + i\eta}, \,\frac{1}{(k - l-k_\gamma)^2 + i\eta},\,\frac{1}{(k-k_g - k_\gamma)^2 + i\eta}, \nonumber \\
    &\,
    \,\frac{1}{k_g\cdot k + i\eta}, \,\frac{1}{n\cdot l + i\eta},\, \frac{1}{n\cdot k-2E_{\gamma}u+ i\eta}\Bigg\} \,.
\end{align}
We note that the naming scheme is chosen to connect them with the corresponding diagrams in Fig.~\ref{fig:antijet_NLO_diagrams_1}. After the reduction, the energy constraint is re-introduced by replacing the last element of each topology, the linear propagator $1/(n\cdot k-2E_\gamma u+i\eta)$, with the delta function\footnote{Strictly speaking, the linear propagator represents only one of the two terms in the reverse-unitarity representation of the delta function. Since the IBP reduction coefficients are independent of the $i\eta$ prescription, the same reduction applies to both terms, while their difference is accounted for when evaluating the cut master integrals.} according to~\eqref{eq:ReverseUnitarity}. The master integrals $G^T_{a_1\ldots a_n}(u)$ given below are therefore understood with this propagator cut. Integrals in which the linear propagator appears with a non-positive index vanish, since the two terms on the right-hand side of~\eqref{eq:ReverseUnitarity} then coincide, or equivalently because of
\begin{equation}
    \label{eq:Cutrelation}
    \left(n\cdot k-2E_{\gamma}u\right)^m \delta(n\cdot k-2E_{\gamma}u)=0\,,\quad m=1,2,\ldots .
\end{equation}
In total, 15 master integrals contribute, 12 of which belong to topology $E$ and 3 to topology $F$. We depict all master integrals in Fig.~\ref{fig:MI_diagrams}.

Most master integrals can be evaluated directly retaining their full $\eps$-dependence. We compute the remaining integrals using the method of differential equations~\cite{Kotikov:1990kg,Kotikov:1991hm,Gehrmann:1999as,Henn:2013pwa}. Since their only non-trivial kinematic dependence is on $u$, the master integrals associated with each topology satisfy a system of first-order linear differential equations,
\begin{equation}
    \partial_u \,\vec{G}^{T}(u) = A_T(u,\eps) \,\vec{G}^{T}(u)\,,
\end{equation}
where the vector $\vec G^T(u)$ also includes the master integrals of sub-topologies, obtained from the corresponding topology by setting one or more propagator powers to zero. We use \texttt{CANONICA}~\cite{Meyer:2017joq} to find a rational change of basis that brings the matrices $A_T(u,\eps)$ into canonical
form~\cite{Meyer:2016slj,Weinzierl:2022eaz}, in which the dependence on $\eps$ factorises. The system can then be solved order by order in $\eps$ in terms of iterated integrals, namely multiple polylogarithms~\cite{goncharov2011multiplepolylogarithmscyclotomymodular}. We determine the integration constants by imposing the regularity of the individual diagrams at the respective endpoints $u\to 0$ or $u\to 1$, and by requiring that their integrals over $u$ reproduce the corresponding integrated amplitudes previously computed in~\cite{Bartocci:2026eeq}. For further details of the calculation, we refer the reader to~\cite{3137480}. The results of all master integrals are collected in an ancillary $\texttt{Mathematica}$ notebook accompanying this publication. We independently validated our analytic results through a numerical evaluation of the master integrals.

The contributions of the individual diagrams to $f_{\text{NLO}}(u)$ are collected in Appendix~\ref{app:u-dep-results}. Summing all contributions yields 
\begin{align}
\label{eq:fNLO_result}
\begin{split}
f_{\text{NLO}}(u)&=2\bigg\{\frac{u}{\eps}(3+2\ln \bar{u})-\bar{u}\Big(4\Li_2(\bar{u})-3 \ln^2 \bar{u}+8\ln \bar{u}-4-\frac{2\pi^2}{3}\Big)-3u\ln u-5\bigg\}C_F\\
&+\frac{1}{2}\bigg\{\frac{8u}{\eps^2}-\frac{4}{\eps}\left[u(3\ln u+4\ln \bar{u})-2(1+u)\right]+8 \bar{u} \Li_2(\bar{u})+(3u\bar{u}-6\bar{u}+10u)\ln^2 \bar{u}\\
&+(10-3\bar{u})u\ln^2u+16 u\ln \bar{u}\ln u-3(2u\bar{u}-\bar{u}+12u)\ln \bar{u}+(6u\bar{u}-12\bar{u}-25u)\ln u\\
&-\Big(43+\frac{4\pi^2}{3}\Big)\bar{u}+67\bigg\}C_A +\mathcal{O}(\eps)\,,
\end{split}
\end{align}
and thus, with $L \equiv \ln\frac{-2E_\gamma\omega_1}{\mu^2}$, we obtain the final expression for the NLO corrections to the bare radiative charm-loop jet function in the limit $m_c \to 0$,
\begin{align}
\label{eq:antijet_NLO_udep}
  \bar{J}_{\rm bare}^{(1)}(u,\omega_1) &=  \frac{\alpha_s}{4\pi}\,\frac{E_\gamma}{2\pi^2\omega_1}\Bigg\{
  C_A\bigg[\frac{2u}{\eps^2}+\frac{u}{\eps}\Big(2-4L-3\ln u-4\ln\bar u\Big)
  + 4uL^2 - uL\big(4-6\ln u-8\ln\bar u\big) \nonumber\\
  &\qquad + 2\bar u\,\Li_2(\bar u)
  + \frac{u(7+3u)}{4}\ln^2 u + \frac{3u\bar u-6\bar u+10u}{4}\ln^2\bar u + 4u\ln u\ln\bar u \nonumber\\
  &\qquad + \frac{6u\bar u+4\bar u-13u}{4}\ln u + \frac{-6u\bar u+15\bar u-20u}{4}\ln\bar u
  + \frac{43}{4}u - \frac{\pi^2}{3}\bar u\bigg] \nonumber\\
  &\quad + C_F\bigg[\frac{u}{\eps}\big(3+2\ln\bar u\big) - 2uL\big(3+2\ln\bar u\big)
  - 4\bar u\,\Li_2(\bar u) + 3\bar u\ln^2\bar u \nonumber\\
  &\qquad - (8\bar u+2u)\ln\bar u - (3u+2\bar u)\ln u + \frac{2\pi^2}{3}\bar u - 4(1+u)\bigg]\Bigg\}
  + \mathcal{O}(\eps)\,.
\end{align}
Integration over $u$ reproduces the result of~\cite{Bartocci:2026eeq}, which is a non-trivial cross-check, since most of the $u$-dependent master integrals were evaluated in closed form, and the $u$-integrated result was used only to fix the boundary constants of two of them.

\subsection{Renormalisation factor}
\label{subsec:singularitystructure}
RG consistency allows us to determine the one-loop renormalisation factor of the radiative charm-loop jet function, defined by
\begin{equation}
\label{eq:ZJbar_def}
\bar{J}_{\rm ren}(u,\omega_1;\mu)=\int_0^1\!dv\int_{-\infty}^{\infty}\!d\omega_1'\,
Z_{\bar J}(u,\omega_1,v,\omega_1';\mu)\,
\bar{J}_{\rm bare}(v,\omega_1')\,, 
\end{equation}
from those of the other functions entering the factorisation theorem~\eqref{eq:factorisationtheorem}. The renormalisation factor $Z_J(p^2,p^{\prime \, 2};\mu)$ of the quark jet function is known~\cite{Becher:2006qw} (see~\eqref{eq:Z-jet}), and some of us previously computed $Z_{17}(\omega,\omega_1,\omega',\omega'_1;\mu)$ for the subleading shape function $g_{17}(\omega,\omega_1;\mu)$~\cite{Bartocci:2024bbf} (see~\eqref{eq:Z-g17}). In this work, we provide the remaining renormalisation factors $Z_{\mathcal{O}_{7\gamma}}(\mu)$ for the electromagnetic dipole operator $\mathcal{O}_{7\gamma}$ and $\mathbf{Z}_{\mathcal{O}_{12}}(u,v;\mu)$ for the current-current operators $\mathcal{O}_{1,2}^q$, given in~\eqref{eq:Z7gamma} and~\eqref{eq:Z12}, respectively. 
We do not present the anomalous dimension or discuss the RG evolution of the jet function, since our strategy is to evolve the hard and soft functions to the hardcollinear scale. Instead, we use $Z_{\bar J}$ to predict the pole structure of the two-loop results, providing a consistency check of the calculation.

We first recall that the Abelian part of $Z_{17}$ (see~\eqref{eq:CFZfactor}) coincides with the renormalisation factor of the leading shape function~\cite{Bosch:2004th}. RG consistency of the leading-power factorisation formula for the $Q_{7\gamma}$-$Q_{7\gamma}$ contribution therefore implies
\begin{align}
&|Z_{\mathcal{O}_{7\gamma}}(\mu)|^{-2} \, m_b\int_{-\infty}^{\bar\Lambda}\!d\tilde\omega\,
Z_J\!\left( m_b (p_+ + \tilde\omega), m_b (p_+ + \omega);\mu\right)
\left[\delta(\tilde{\omega} - \omega') + Z_{17}^{(1)}(\tilde\omega,\omega';\mu)\Big\vert_{C_F}\right] \nonumber \\
= \, &\delta(\omega-\omega')
+\mathcal{O}(\alpha_s^2)\,, 
\end{align}
where $p_+ = m_b - 2E_\gamma$. The one-loop renormalisation factor of the radiative charm-loop jet function can thus be determined from $Z_{\mathcal{O}_{7\gamma}}$, the $(1,1)$ entry of the matrix $\mathbf{Z}_{\mathcal{O}_{12}}$ and the non-Abelian part of $Z_{17}$ given in~\eqref{eq:CAZfactor}, 
\begin{align}
\label{eq:ZJbar}
&Z^{(1)}_{\bar J}(u,\omega_1,v,\omega'_1;\mu) \nonumber \\
=\, &
-\delta(u-v)\,\delta(\omega_1-\omega'_1)\,
Z^{(1)}_{\mathcal{O}_{7\gamma}}(\mu)
+\delta(\omega_1-\omega'_1)\,
\big[\mathbf{Z}^{(1)}_{\mathcal{O}_{12}}\big]_{11}(u,v;\mu)
-\delta(u-v)\,
Z_{17}^{(1)}(\omega'_1,\omega_1;\mu)\Big|_{C_A}
\nonumber\\
= \, &-\frac{\alpha_s}{4\pi}\Bigg\{
C_A\,\delta(u-v)\,\delta(\omega_1-\omega'_1)
\left[
\frac{2}{\eps^2}
-\frac{2}{\eps}\ln\frac{-m_b \omega_1}{\mu^2}
-\frac{1}{\eps}\ln(u\bar u)
+\frac{3}{2\eps}
\right] \\
&+\delta(\omega_1-\omega'_1)\,
\frac{2C_F-C_A}{\eps}\,V_{\text{ERBL}}(u,v)
\nonumber\\
&+\delta(u-v)\,\frac{2C_A}{\eps}
\Bigg[
\theta(\omega_1)\theta(\omega'_1)
\left(
\frac{\omega'_1}{\omega_1^2}\,
\theta(\omega_1-\omega'_1)
-F(\omega'_1,\omega_1)
\right)
+ (\omega_1,\omega'_1)
\to(-\omega_1,-\omega'_1) 
\Bigg] \Bigg\} + \mathcal{O}(\alpha_s^2)\,.  \nonumber
\end{align}
Here $F(\omega_1,\omega'_1)$ is the Lange--Neubert
kernel~\cite{Lange:2003ff},
\begin{equation}
F(\omega_1,\omega'_1) = \omega_1 \left[
\frac{\theta(\omega'_1-\omega_1)}{\omega'_1(\omega'_1-\omega_1)}
+ \frac{\theta(\omega_1-\omega'_1)}{\omega_1(\omega_1-\omega'_1)}
\right]_{+}^{(\omega_1)},
\qquad \omega_1,\omega'_1 > 0 \,, 
\end{equation}
written with the plus prescription acting on its first argument. Consequently, the plus prescription in $F(\omega'_1,\omega_1)$ appearing in~\eqref{eq:ZJbar} acts on the integration variable $\omega'_1$. The second equality in~\eqref{eq:ZJbar} is understood
under the convolution in~\eqref{eq:ZJbar_def}. To obtain this representation, we use the analyticity of the jet function in the upper half of the complex $\omega'_1$ plane to reduce the generalised plus distributions appearing in $Z_{17}$, see also
the discussion in~\cite{Beneke:2022msp,Bartocci:2024bbf}.

When convoluted with the massless LO bare jet function~\eqref{eq:antijet_LO_udep}, the last square bracket in~\eqref{eq:ZJbar} acts as the identity up to $\mathcal{O}(\eps)$ corrections. Since it multiplies a single pole, these $\mathcal{O}(\eps)$ corrections contribute only finite terms. For the extraction
of the NLO poles, we may therefore write
\begin{align}
Z_{\bar J}^{(1)}(u,\omega_1,v,\omega'_1;\mu) \to 
-\frac{\alpha_s}{4\pi}\,\delta(\omega_1-\omega'_1)
\Bigg\{& C_A\,\delta(u-v)
\left[ \frac{2}{\eps^2} -\frac{2}{\eps}\ln\frac{-m_b \omega_1}{\mu^2} -\frac{1}{\eps}\ln(u\bar u)
+\frac{7}{2\eps} \right] \nonumber\\
&+\frac{2C_F-C_A}{\eps}\,V_{\mathrm{ERBL}}(u,v)
\Bigg\}\,. 
\end{align}
Since the ERBL kernel multiplies a single pole, only the $\mathcal{O}(\eps^0)$ term of the LO jet function, which is proportional to $v$, is needed to extract the NLO poles.
The remaining convolution reduces to
\begin{equation}
    \int_0^1 \! dv \, V_{\rm ERBL}(u,v)\,v
    = u\ln\bar u + \frac{3}{2}u \,. 
\end{equation}
We therefore obtain for the NLO singularities 
\begin{align}
\label{eq:antijet_poles_NLO}
&\int_0^1\!dv\int_{-\infty}^{\infty}\!d\omega_1'\,
Z^{(1)}_{\bar J}(u,\omega_1,v,\omega_1';\mu)\,
\bar{J}^{(0)}_{\rm bare}(v,\omega_1')
\\
&= -\frac{\alpha_s}{4\pi}\,
\frac{E_\gamma}{2\pi^2 \omega_1} \Bigg\{ C_A\left[ \frac{2u}{\eps^2} + \frac{u}{\eps} \left( 2 - 4\ln\frac{-2E_\gamma \omega_1}{\mu^2} -3\ln u-4\ln\bar u \right) \right] + \frac{C_F u}{\eps}\left(3+2\ln\bar u\right) \Bigg\}
+\mathcal{O}(\eps^0)\,,  \nonumber 
\end{align}
with $2E_\gamma = m_b$ up to power corrections. The $\mathcal{O}(\eps)$ term of $\bar{J}^{(0)}_{\rm bare}$ must be retained because it multiplies the double pole and therefore contributes to the single-pole term. The result cancels the singularities of the explicit two-loop calculation~\eqref{eq:antijet_NLO_udep}, as required
by~\eqref{eq:ZJbar_def},
\begin{equation}
\bar J_{\rm bare}^{(1)}
+Z_{\bar J}^{(1)}\otimes\bar J_{\rm bare}^{(0)}
=\mathcal{O}(\eps^0)\,.
\end{equation}
This cancellation at fixed $u$ provides a non-trivial cross-check of the factorisation framework and our calculation. Upon integration over $u$, the $C_F$ contribution vanishes because
$\int_0^1 du\,u(3+2\ln\bar u)=0$. We then obtain the counterterm corresponding to the integrated two-loop result given in~\cite{Bartocci:2026eeq},
\begin{align}
\label{eq:antijet_poles_2_NLO}
\int_0^1 \! du \, dv\int_{-\infty}^{\infty}\!d\omega_1'\,
Z^{(1)}_{\bar J}(u,\omega_1,v,\omega_1';\mu)\,
\bar{J}^{(0)}_{\rm bare}(v,\omega_1')
= \frac{\alpha_s C_A}{4\pi}\,
\frac{E_\gamma}{2\pi^2 \omega_1} \left[ \frac{-1}{\eps^2} + \frac{2}{\eps} \ln\frac{-2E_\gamma \omega_1}{\mu^2} -\frac{19}{4\eps} \right]
+\mathcal{O}(\eps^0)\,.
\end{align}

Determining the complete renormalisation kernel~\eqref{eq:ZJbar} constitutes a non-trivial extension of the analysis in~\cite{Bartocci:2026eeq}, as it fixes the pole structure with full dependence on the momentum fraction $u$. Furthermore, the same kernel
also allows us to predict the NLO poles for $m_c\neq0$ by convolving it with the LO jet function including its 
charm-mass dependence and adding the known contribution from charm-mass renormalisation. These predictions provide a further cross-check of the massive two-loop calculation, which will be presented elsewhere. Finally, using~\eqref{eq:ZJbar_def}, we obtain the renormalised radiative charm-loop jet function at NLO,
\begingroup
\interdisplaylinepenalty=10000
\begin{align}
\bar{J}(u,\omega_1;\mu)
= \, &\frac{E_\gamma}{2\pi^2\omega_1}
\Biggl\{u+\frac{\alpha_s(\mu)}{4\pi}\Biggl[
\notag\\
\frac{C_A}{4} &\biggl(
4uL^2+2uL\bigl(2\ln(u\bar u)-7\bigr)
+u(2-3\bar u)\ln^2u
+u(2+3\bar u)\ln^2\bar u +u(6\bar u-5)\ln u
\notag\\
&\quad-(6u\bar u+10u+\bar u)\ln\bar u
+27u-\frac{2\pi^2u}{3}
\biggr)
\notag\\[2pt]
+ \left(C_F-\frac{C_A}{2}\right) &\biggl(
-uL(3+2\ln\bar u)
+u\ln^2\frac{u}{\bar u}
+(1+2\bar u)\ln^2\bar u -2(1+\bar u)\operatorname{Li}_2(\bar u) -2\ln u+(u-6\bar u)\ln\bar u
\notag\\
&\quad
+\frac{2\pi^2}{3}\bar u-4(1+u)
\biggr)
\Biggr]\Biggr\}
+\mathcal{O}(\alpha_s^2)\,.
\label{eq:antijet_ren_NLO}
\end{align}
This expression differs from the finite part of the bare
expression in~\eqref{eq:antijet_NLO_udep}, because the
$\mathcal{O}(\eps)$ and $\mathcal{O}(\eps^2)$ terms of the bare LO jet function~\eqref{eq:antijet_LO_udep} combine with the poles of $Z_{\bar J}^{(1)}$ to finite contributions.
\endgroup

\section{Conclusions}
\label{sec:conclusion}
In this work, we have investigated the inclusive radiative decay $\bar{B}\to X_s\gamma$ at large photon energies, focusing on the subleading-power resolved-photon contribution arising from the interference between the current-current operators $Q_{1,2}^q$ and the electromagnetic dipole operator $Q_{7\gamma}$. Motivated by the need to reduce the substantial perturbative uncertainties associated with this contribution, we have carried out a systematic analysis of radiative corrections at next-to-leading order in $\alpha_s$. Using a factorisation theorem in soft-collinear effective theory, we have determined all previously unknown perturbative ingredients required for a renormalisation-group improved treatment in the limit of a massless charm quark.

\begingroup \emergencystretch=2em
Using the power counting $m_c^2 \sim \Lambda_{\rm QCD} m_b$, we have computed the $m_c$-independent one-loop hard matching of the weak effective operators onto their SCET counterparts and derived the renormalisation-group equations for the corresponding hard functions. For the non-local evolution of $H_1^q(u)$, we have obtained an approximate analytic solution based on a large-$N_c$ expansion, including the first correction from the ERBL kernel, which enters with a colour-suppressed coefficient. The approximate analytic solution reproduces the numerical solution of the full evolution equation at the per-mille level and is therefore expected to be sufficiently accurate for phenomenological applications. \par \endgroup

Furthermore, we have computed the two-loop radiative charm-loop jet function $\bar J(u,\omega_1;\mu)$ analytically in the limit $m_c\to0$, retaining its full dependence on the fraction $u$ of the large photon momentum carried by the charm quark. We have also derived the complete one-loop renormalisation kernel from renormalisation-group consistency and verified that it reproduces the two-loop pole structure at fixed $u$, providing a non-trivial cross-check of our analysis. Together with the known contribution from charm-mass renormalisation, this kernel also predicts the corresponding singularities for $m_c\neq0$.

Combined with the known radiative corrections to the leading-power quark jet function and the one-loop evolution kernel of the subleading shape function $g_{17}(\omega,\omega_1;\mu)$, recently computed by some of us~\cite{Bartocci:2024bbf}, our results provide the complete theoretical framework for a renormalisation-group improved treatment of this subleading-power resolved-photon contribution. A phenomenological analysis requires the inclusion of charm-quark mass effects. In our formalism, $m_c$ enters only the anti-hardcollinear radiative charm-loop jet function, whose two-loop calculation at finite charm mass therefore remains the only missing perturbative ingredient for a phenomenological analysis at next-to-leading order. The results presented here serve as an important analytic benchmark for this calculation.

\section*{Acknowledgments} 
 We thank Michael Benzke, Christoph Greub and Vladyslav Shtabovenko for useful discussions. This work is supported by  the  Cluster  of  Excellence  ``Precision  Physics,  Fundamental Interactions, and Structure of Matter" (PRISMA$^{++}$ EXC 2118/2) funded by the German Research Foundation (DFG) within the German Excellence Strategy (Project ID 390831469) and also by the DFG Grant HU 901/2-1. The research of P.B. was funded by the European Union's Horizon 2020 Research and Innovation Program under the Marie Sk\l{}odowska-Curie grant
agreement No.101146976.

\begin{appendix}
\section{Collection of NLO results known in the literature}
\label{app:literature}
The renormalisation properties of two functions entering the factorisation formula~\eqref{eq:factorisationtheorem} have already been studied in the literature. The hardcollinear jet function $J$ was computed at NLO and NNLO in~\cite{Becher:2006qw} and subsequently extended to N$^3$LO in~\cite{Bruser:2018rad}. The non-perturbative shape function $g_{17}$ was investigated in our previous work~\cite{Bartocci:2024bbf}, where we derived its one-loop anomalous dimension and obtained closed-form analytic solutions to the corresponding renormalisation-group equation. In this appendix, we collect the results relevant for the present analysis, focusing on the renormalisation factors required for the cancellation of singularities discussed in Section~\ref{subsec:singularitystructure}.

\subsection{Hardcollinear jet function}
\label{subsec:quarkjetfunction}
The hardcollinear quark propagator is defined by
\begin{equation}\label{eq:Jscet}
   \frac{\slashed{n}}{2}\,\bar{n}\cdot p\,{\cal J}(p^2;\mu)
   = \int d^4x\,e^{-ip\cdot x}\,\langle 0|T\{
   \chi_{hc}(0)\,\overline\chi_{hc}(x)\}|0\rangle \,, 
\end{equation} 
where $\chi_{hc}$ denotes the gauge-invariant massless strange-quark field. The corresponding hardcollinear jet function $J$ is defined through the discontinuity of the propagator,
\begin{equation}
   J(p^2;\mu) 
   =  \frac{1}{2 i \pi}\, {\rm Disc}\left[ i{\cal J}(p^2;\mu)    \right] \,. 
\end{equation}
In the $\overline{\text{MS}}$ scheme, the renormalised jet function is related to its bare counterpart by
\begin{equation}
    J_{\rm{ren}}(p^2;\mu) = \int dp^{\prime \, 2} \, Z_J(p^2,p^{\prime \, 2};\mu) \, J_{\rm bare}(p^{\prime \, 2}) \,. 
\end{equation}
The resulting renormalised jet function at one-loop order reads~\cite{Bosch:2004th}
\begin{align}
\label{eq:Jren}
   J_{\rm ren}(p^2;\mu)
   =\delta(p^2)+ \, &\frac{\alpha_s C_F}{4\pi} \Bigg[
   \left(7 - \pi^2 \right)
   \delta(p^2)
   + 4 \left( \frac{\ln{p^2/\mu^2}}{p^2} \right)_{\!*}^{\![\mu^2]}
   - 3
   \left( \frac{1}{p^2} \right)_{\!*}^{\![\mu^2]} \Bigg] +\mathcal{O}(\alpha_s^2)\,,
\end{align}
while the corresponding renormalisation kernel is given by~\cite{Becher:2006qw}
\begin{equation}
\label{eq:Z-jet}
   Z_J(p^2,p^{\prime \, 2};\mu) = \delta(p^2 - p^{\prime \, 2}) - \frac{\alpha_s C_F}{4\pi} \left[ \delta(p^2 - p^{\prime \, 2}) \left(  \frac{4}{\eps^2} + \frac{3}{\eps} \right) - \frac{4}{\eps} \left( \frac{1}{p^2 - p^{\prime \, 2}} \right)_{\!*}^{\![\mu^2]} \right] + \, \mathcal{O}(\alpha_s^2) \,. 
\end{equation}
The star distributions appearing in the above expressions are defined as in~\cite{Bosch:2004th}.

\subsection{Subleading shape function \texorpdfstring{$g_{17}$}{g17}}
\label{subsec:g17}
The subleading shape function $g_{17}$ is defined through the Fourier transform of the forward matrix element between static $\bar{B}$-meson states,
\begin{align}\label{eq:g17def}
   g_{17}(\omega,\omega_1;\mu) = \frac{1}{2M_B} \int\frac{dr}{2\pi}\,e^{-i\omega_1 r}
    \int\frac{dt}{2\pi}\,e^{-i\omega t} \, \langle\bar B_v| \mathcal{O}_{17}(t,r) |\bar B_v\rangle \,, 
\end{align}
where $\mathcal{O}_{17}$ denotes the non-local operator in Heavy-Quark Effective Theory (HQET)~\cite{Benzke:2010js,Bartocci:2024bbf}
\begin{equation}
\label{eq:Q17}
    \mathcal{O}_{17}(t,r) = \big(\bar h_v S_n\big)_-(tn)\,
    \nbslash \big(S_n^\dagger S_{\bar n}\big)_+(0)\,
    i\gamma_\alpha^\perp\bar n_\beta\,
    \big(S_{\bar n}^\dagger\,g_s G_s^{\alpha\beta} S_{\bar n} 
    \big)_+(r\bar n)\,
    \big(S_{\bar n}^\dagger h_v\big)_+(0) \,. 
\end{equation}
The soft Wilson lines $S_n$ arise from the decoupling of the hardcollinear strange-quark propagator, while $S_{\bar n}$ originate from the decoupling of the anti-hardcollinear propagators in the quark loop. The subscripts ``$+$'' and ``$-$'' refer to the two branches of the Keldysh contour~\cite{Schwinger:1960qe,Keldysh:1964ud} (see also~\cite{Becher:2007ty} for a concise summary). Fields labelled by ``$+$'' belong to the amplitude and are evaluated using the standard HQET Feynman rules for time-ordered products, whereas fields labelled by ``$-$'' belong to the complex-conjugate amplitude and are evaluated using complex-conjugate Feynman rules for anti-time-ordered products.

The one-loop renormalisation factor $Z_{17}$ of the ``di-light-cone operator'' $\mathcal{O}_{17}$ was derived in~\cite{Bartocci:2024bbf}. It can be decomposed into Abelian and non-Abelian contributions as
\begin{equation}
\label{eq:Z-g17}
    Z_{17}(\omega,\omega_1,\omega',\omega'_1;\mu) = \delta(\omega - \omega') \delta(\omega_1 - \omega'_1) + \delta(\omega_1 - \omega'_1)Z_{17}^{(1)}\Big\vert_{C_F} + \delta(\omega - \omega')Z_{17}^{(1)}\Big\vert_{C_A} + \mathcal{O}(\alpha_s^2) \,. 
\end{equation}
The Abelian contribution coincides with the renormalisation factor of the leading shape function~\cite{Bosch:2004th} and is given by
\begin{align}
\label{eq:CFZfactor}
    Z_{17}^{(1)}\Big\vert_{C_F}  = 
     \frac{\alpha_s C_F}{4\pi}  \bigg\{ \left(\frac{2}{\eps^2} + \frac{4}{\eps} \ln \frac{\mu}{\bar{\Lambda}-\omega} - \frac{2}{\eps} \right) \delta(\omega-\omega')
    - \frac{4}{\eps} \left[ 
\frac{\theta(\omega'-\omega)}{\omega'-\omega} \right]_+^{(\bar{\Lambda})} \bigg\} \,, 
\end{align}
where $\bar{\Lambda} = M_B - m_b$ denotes the difference between the $B$-meson mass and the $b$-quark pole mass. The non-Abelian contribution takes the form
\begin{align}
\label{eq:CAZfactor}
    Z_{17}^{(1)}\Big\vert_{C_A} = \, &\frac{\alpha_s C_A}{4\pi}  
    \bigg\{ \left( \frac{1}{\eps^2} + \frac{1}{\eps}\ln \frac{\mu^2}{\omega_1^2}\right) \delta(\omega_1-\omega'_1) \nonumber \\[0.2em]
    + \, &\frac{2}{\eps} \frac{\omega_1}{(\omega'_1)^2} \big[ \theta(\omega_1)\theta(\omega'_1 - \omega_1) - \theta(-\omega_1) \theta(\omega_1 - \omega'_1) \big] - \frac{1}{\eps} \, \text{Re} \, [H_+(\omega_1,\omega'_1)+H_-(\omega_1,\omega'_1)] \bigg\} \,.
\end{align}
The plus distribution $[\ldots]_+^{(\bar\Lambda)}$ is defined as in~\cite{Bosch:2004th,Bartocci:2024bbf}, whereas the distributions $H_\pm$ are as introduced in~\cite{Beneke:2022msp}.

The structure of the renormalisation factor implies that the anomalous dimension decomposes into two contributions, each acting on only one of the light-cone variables,
\begin{align}
\label{eq:gamma17}
 \gamma_{17}(\omega,\omega_1,\omega',\omega'_1;\mu)
 = \frac{\alpha_s}{\pi} \Big\{ C_F \, \delta(\omega_1 - \omega'_1) \gamma_{n}(\omega,\omega';\mu)
+ \frac{C_A}{2} \, \delta(\omega - \omega') \gamma_{\bar{n}}(\omega_1,\omega'_1;\mu) \Big\} \,.  
\end{align}
Consequently, the renormalisation-group evolution factorises into independent contributions associated with the two light-cone directions. The anomalous dimension governing the evolution in the $n$ direction is given by
\begin{equation}
\label{eq:gamman}
\gamma_{n}(\omega,\omega';\mu) =  \left( 2 \ln \frac{\mu}{\bar{\Lambda}-\omega} - 1 \right) \delta(\omega-\omega')
    - 2 \left[ 
\frac{\theta(\omega'-\omega)}{\omega'-\omega} \right]_+^{(\bar{\Lambda})}\,,
\end{equation}
while the corresponding anomalous dimension in the $\bar n$ direction reads
\begin{align}
\label{eq:gammanbar}
    \gamma_{\bar{n}}(\omega_1,\omega'_1;\mu) &=\,  \ln \frac{\mu^2}{\omega_1^2} \, \delta(\omega_1-\omega'_1)\nonumber \\
    &\,  - \text{Re} \, [H_+(\omega_1,\omega'_1)+H_-(\omega_1,\omega'_1)]+ \frac{2 \omega_1}{(\omega'_1)^2} \big[ \theta(\omega_1)\theta(\omega'_1 - \omega_1) - \theta(-\omega_1) \theta(\omega_1 - \omega'_1) \big]\,.
\end{align}
Analytic closed-form solutions to the corresponding evolution equations were derived in~\cite{Bartocci:2024bbf}.

\section{Renormalisation factors of WET operators and evanescent contributions}
\label{app:evanescent_ops}
In this appendix, we collect the renormalisation factors of the weak effective theory (WET) operators and discuss the contributions of evanescent operators to the hard matching functions presented in Section~\ref{sec:hardmatching}. Throughout, we distinguish ultraviolet (UV) and infrared (IR) divergences, associating the former with the WET renormalisation scale $\nu$ and the latter with the factorisation scale $\mu$.

For the electromagnetic dipole operator $Q_{7\gamma}$, the UV-renormalised $b\to s\gamma$ on-shell matrix element is obtained from
\begin{align}
\left\langle{Q_{7\gamma}}\right\rangle_{\rm ren} =
Z_{77}^{\overline{\rm MS}}\,
Z_{m_b}^{\overline{\rm MS}}\,
\left\langle{Q_{7\gamma}}\right\rangle_{\rm bare}\,. 
\end{align}
Here, $\langle Q_{7\gamma}\rangle_{\rm bare}$, given in~\eqref{eq:H7poles}, denotes the matrix element of the bare operator evaluated with on-shell renormalised external fields. It therefore already includes the wave-function renormalisation factors $Z_{2,b}^{1/2,\rm{OS}},Z_{2,s}^{1/2,\rm{OS}}$, which at one-loop order are given by
\begin{align}
\label{eq:Z2b}
Z_{2,b}^{\rm OS} &= 1 - \frac{\alpha_s C_F}{4\pi}\Big( \frac{1}{\eps_{\UV}} + \ln\frac{\nu^2}{m_b^2}
 + \frac{2}{\eps_{\IR}} + 2\ln\frac{\mu^2}{m_b^2} + 4 \Big) + \mathcal{O}(\alpha_s^2)\,, \\
\label{eq:Z2s}
Z_{2,s}^{\rm OS} &= 1 - \frac{\alpha_s C_F}{4\pi}\Big( \frac{1}{\eps_{\UV}}
 - \frac{1}{\eps_{\IR}} + \ln\frac{\nu^2}{\mu^2} \Big) + \mathcal{O}(\alpha_s^2)\,. 
\end{align}
The self-energy correction of the massless strange quark is scaleless and therefore vanishes in dimensional regularisation when UV and IR divergences are identified. However, since we keep these divergences separate, the corresponding wave-function renormalisation factor $Z_{2,s}^{\rm OS}$ must be retained to account for its individual UV and IR contributions. The operator and mass renormalisation factors are defined in the $\overline{\rm MS}$ scheme and read
\begin{align}
Z_{77}^{\overline{\rm MS}} &= 1 + \frac{\alpha_s C_F}{4\pi}\,\frac{4}{\eps_{\UV}} + \mathcal{O}(\alpha_s^2)\,, \\
Z_{m_b}^{\overline{\rm MS}} &= 1 - \frac{\alpha_s C_F}{4\pi}\,\frac{3}{\eps_{\UV}} + \mathcal{O}(\alpha_s^2)\,.
\end{align}
Here, $Z_{m_b}^{\overline{\rm MS}}$ accounts for the renormalisation of the $b$-quark mass appearing in the definition of $Q_{7\gamma}$, such that the renormalised operator is expressed in terms of the $\overline{\rm MS}$ mass $\overline{m}_b(\nu)$. The one-loop vertex correction is UV finite, and the remaining UV pole in~\eqref{eq:H7poles} originates from the external-field renormalisation factors~\eqref{eq:Z2b} and~\eqref{eq:Z2s}. This divergence is cancelled by the product of the operator and mass renormalisation factors,
\begin{equation}
Z_{77}^{\overline{\rm MS}}\, Z_{m_b}^{\overline{\rm MS}}
= 1 + \frac{\alpha_s C_F}{4\pi}\,\frac{1}{\eps_{\UV}} + \mathcal{O}(\alpha_s^2)\,, 
\end{equation}
which coincides with the one-loop renormalisation factor of the tensor current.

The renormalisation of the four-quark operators $Q_1^q$ and $Q_2^q$ requires an extension of the physical operator basis by two evanescent operators at one-loop order in QCD~\cite{Beneke:2009ek}. In the conventions adopted here, these are defined as
\begin{align}
\label{eq:E1}E_1^q &= (\bar s_j \gamma_\mu \gamma_\nu \gamma_\rho (1-\gamma_5) q_j)(\bar q_i \gamma^\mu \gamma^\nu \gamma^\rho (1-\gamma_5) b_i) - 16\, Q_1^q\,, \\
\label{eq:E2}E_2^q &= (\bar s_j \gamma_\mu \gamma_\nu \gamma_\rho (1-\gamma_5) q_i)(\bar q_i \gamma^\mu \gamma^\nu \gamma^\rho (1-\gamma_5) b_j) - 16\, Q_2^q\,,
\end{align}
and vanish identically in four dimensions but are non-zero in dimensional regularisation. Although their matrix elements are of $\mathcal{O}(\eps)$, they generate finite contributions to the hard functions when multiplied by the $1/\eps_{\UV}$ poles of the renormalisation constants. The renormalisation matrix is expanded as $Z_{ij}=\delta_{ij}+Z^{(1)}_{ij}+\mathcal{O}(\alpha_s^2)$, with the one-loop contribution
\begin{equation}
\label{eq:Zij}
Z_{ij}^{(1)} = \frac{\alpha_s}{4\pi}\frac{1}{\eps_{\UV}} \left( \begin{array}{cccc}
\; 6 C_F-3 C_A \; & \; 3 \; & \; C_F- \frac12 C_A \; & \; \frac{1}{2} \; \\
\; 3 \; & \; 6 C_F-3 C_A \; & \; \frac{1}{4} \; & \; C_F- \frac14 C_A \; 
\end{array} \right),
\end{equation}
where $i=1,2$ and the column index $j$ refers to the basis $(Q_1^q,Q_2^q,Q_3^q,Q_4^q)\equiv(Q_1^q,Q_2^q,E_1^q,E_2^q)$. The renormalisation matrix agrees with the result of~\cite{Beneke:2009ek} after transforming to the operator basis adopted here. The UV-renormalised $b\to s\bar q q$ on-shell matrix elements of the physical operators, evaluated with on-shell external partons, are then obtained from
\begin{align}
\left\langle{Q_i^q}\right\rangle_{\rm ren} = \sum_{j=1}^{4} Z_{ij} \, \langle{Q_j^{q}}\rangle_{\rm bare}\,, 
\end{align}
where the bare matrix elements in~\eqref{eq:H1_bare_full} are again computed with on-shell renormalised fields, i.e.\ they include the factors $Z_{2,b}^{1/2,\rm{OS}}\,Z_{2,s}^{1/2,\rm{OS}}\,Z_{2,q}^{\rm OS}$, with $Z_{2,q}^{\rm OS}=Z_{2,s}^{\rm OS}$ in~\eqref{eq:Z2s} for the two external light quarks. Note that we assume all partons except the $b$ quark to be massless. The first two columns of~\eqref{eq:Zij} cancel the UV divergences in~\eqref{eq:H1_bare_full}. The remaining two columns account for the evanescent operator contributions, which generate finite terms through the multiplication of the $1/\eps_{\UV}$ poles of the renormalisation constants with the $\mathcal{O}(\eps)$ tree-level matrix elements of $E_{1,2}^q$. The UV poles associated with external-field renormalisation also generate finite terms when multiplied by
$\mathcal{O}(\eps)$ evanescent structures. These contributions
are included in our calculation. An additional finite conversion is required because the hard matching functions are computed using the evanescent Fierz-rearranged operators $E_1^q$ and $E_2^q$. However, in this paper, we adopt the evanescent prescription of~\cite{Bell:2009nk}, with subtraction terms $-(16-4\eps)Q_i^q$, whereas the prescription
in~\eqref{eq:E1} and~\eqref{eq:E2} uses $-16Q_i^q$.
The $\mathcal{O}(\eps)$ difference between these prescriptions
induces a finite renormalisation through its multiplication
by UV poles. Expressing the result in the operator basis
and evanescent prescription adopted here therefore requires
the finite one-loop conversion matrix
\begin{equation}
    \Delta_B =
    \frac{\alpha_s}{4\pi}
    \begin{pmatrix}
        C_A & -1 \\
        1 & -C_A
    \end{pmatrix}.
\end{equation}

\section{Details on the two-loop radiative charm-loop jet function}
\label{app:u-dep-results}
In this appendix we collect the $u$-dependent results for all the diagrams shown in
Figures~\ref{fig:antijet_NLO_diagrams_1} and~\ref{fig:antijet_NLO_diagrams_2}.
For each diagram we give the corresponding NLO function of $u$, including its colour factor, to be
inserted in~\eqref{eq:two_loop_massless_udep}, such that their sum reproduces
$f_{\text{NLO}}(u)$ in~\eqref{eq:fNLO_result}. The results
read\footnote{By (G) we denote the sum of diagrams (G.1) and (G.2), and by
(I) the sum of diagrams (I.1) and (I.2).}:
\begin{align}
    \begin{split}
        f_{\text{A}}(u)=\,C_F\Bigg\{\frac{u}{\eps}\left[\bar{u}\ln\frac{u}{\bar{u}}-4\right]+\frac{1}{2}\bigg[12 \bar{u}-3u \ln (\bar{u}u) \left(\bar{u} \ln\frac{u}{\bar{u}}-4\right)-20\bigg]+\mathcal{O}(\eps)\Bigg\}\,,
    \end{split}
\end{align}
\begin{align}
    \begin{split}
        f_{\text{B}}(u)=\,\Big(C_F-\frac{1}{2}C_A\Big)\Bigg\{\frac{u \bar{u}}{\eps}\ln\frac{\bar{u}}{u}+\frac{u}{6}\bigg[9\, \bar{u} \ln\frac{u}{\bar{u}}\, \ln (u\bar{u})+4 \left(\pi ^2-6\right)\bigg]+\mathcal{O}(\eps)\Bigg\}\,,
    \end{split}
\end{align}
\begin{align}
    \begin{split}
        f_{\text{C}}(u)&=\,C_A\Bigg\{\frac{u}{2\eps}\left(3\,\bar{u}\ln\frac{\bar{u}}{u}+4+\frac{\pi^2}{3}\right)-\frac{9}{4}u\bar u
\left(\ln^2 \bar{u}-\ln^2 u\right)\\&
-u\left[\Big(\frac{\pi^2}{6}(2+\bar{u})+1+3u\Big)\ln \bar{u}+\Big(\frac{\pi^2}{6}u+2+3\bar{u}\Big)\ln u-\zeta_3\right]-\frac{\pi^2}{6}(u-\bar{u})-6\bar{u}+8+\mathcal{O}(\eps)\Bigg\}\,,
    \end{split}
\end{align}
\begin{align}
    \begin{split}
        f_{\text{D}}(u)&=\,C_F\Bigg\{\frac{\bar{u}}{\eps}\left(\bar{u}\ln\frac{u}{\bar{u}}-\ln(\bar{u}^3u)\right)
        \\& +\frac{3}{2}u\bar u\,\ln^2 u+\frac{3}{2}\bar{u}(\bar{u}+3)\ln^2 \bar{u}+4\bar{u}\,\ln u\,\ln\bar u-8u\,\ln u-2(1+6\bar{u})\ln \bar{u}-6 u+\mathcal{O}(\eps)\Bigg\}\,,
    \end{split}
\end{align}
\begin{align}
    \begin{split}
        f_{\text{E}}(u)&=\,\Big(C_F-\frac{1}{2}C_A\Big)\Bigg\{\frac{1}{\eps}\left[u\left(\bar{u}\ln\frac{u}{\bar{u}}+10\right)+4\ln \bar{u}\right]-8\bar{u}\,\Li_2(\bar{u})-\frac{3}{2}u\bar u\left(\ln^2 u-\ln^2 \bar{u}\right)\\&-4\bar{u}\,\ln u\,\ln\bar u-4 u\,\ln u-2(1+u)\ln \bar{u}-\frac{2\pi^2}{3}(1-3\bar{u})-8\bar{u}+10+\mathcal{O}(\eps)\Bigg\}\,,
    \end{split}
\end{align}
\begin{align}
    \begin{split}
        f_{\text{F}}(u)&=C_A\Bigg\{\frac{1}{2\eps}\left[u\Big(\bar{u}\ln u-\frac{\pi^2}{3}\Big)+\bar{u}(3+\bar{u})\ln \bar{u}\right]-\frac{3}{4}\bar{u}(1+3\bar{u})\ln^2 \bar{u}-\frac{9}{4}u\bar u\,\ln^2 u
        \\&-2\bar{u}\,\ln u\,\ln\bar u+\left[\frac{\pi^2}{6}u(\bar{u}+2)+3\bar{u}^2+\frac{7}{2}\bar{u}\right]\ln \bar{u}+\left[\frac{\pi^2}{6}u^2+\frac{3}{2}u(2\bar{u}+3)\right]\ln u\\&-u\left(\zeta_3-\frac{1}{2}\right)-\frac{\pi^2}{6}(3u+\bar{u})+\mathcal{O}(\eps)\Bigg\}\,,
    \end{split}
\end{align}
\begin{align}
    \begin{split}
        f_{\text{G}}(u)&=\,C_A\Bigg\{\frac{3u}{\eps^2}-\frac{1}{\eps}\bigg[u\big[(4+u)\ln u+(4+\bar{u})\ln \bar{u}\big]+5\bar{u}-8\bigg]+\frac{3}{2}u\,\Big((2+\bar{u})\ln^2 \bar{u}\\
        &+(2+u)\ln^2 u\Big)+6u\,\ln u\,\ln\bar u-\left(5+8u-3u^2\right)\ln \bar{u}-\left(4+7u+3 u^2\right)\ln u\\
        &+\frac{\pi^2}{2}u-12\bar{u}+20+\mathcal{O}(\eps)\Bigg\}\,,
    \end{split}
\end{align}
\begin{align}
    \begin{split}
    f_{\text{H}}(u)&=\,C_A\Bigg\{-\frac{u}{\eps^2}-\frac{1}{\eps}\bigg[\bar{u}^2 \ln\frac{u}{\bar{u}}-\ln u-(2u-\bar{u}) \ln \bar{u}+1+3u\bigg]-\frac{u}{2}\Big[(4-3u)\ln^2u\\&+(1+3u)\ln^2\bar{u}+4\ln u\,\ln\bar u\Big]-(3u+1)\Big[(u-2)\ln u-(1+u)\ln \bar{u}\Big]-\frac{\pi^2}{6}u-8u-4+\mathcal{O}(\eps)\Bigg\}\,,
    \end{split}
\end{align}
\begin{align}
    \begin{split}
        f_{\text{I}}(u)&=\,C_A\Bigg\{\frac{2u}{\eps^2}+\frac{1}{\eps}\bigg[\bar{u}^2 \ln\frac{u}{\bar{u}}-(1+u) \ln u-(4 u-\bar{u}) \ln \bar{u}+7u+2\bar{u}\bigg]\\&
        +\frac{u}{2}\Big[(5-3 u)\ln^2 u+(5+3 u)\ln^2 \bar{u}+8\ln u\,\ln\bar u\Big]-(4+6 u-3 u^2)\ln u-(2+9 u+3 u^2)\ln \bar{u}\\
        &+\frac{\pi^2}{3} u+13 u+7+\mathcal{O}(\eps)\Bigg\}\,.
    \end{split}
\end{align}
The poles proportional to $\pi^2$ in $f_{\text{C}}(u)$ and $f_{\text{F}}(u)$, as well as the terms proportional to $\zeta_3$, cancel in their sum.

\end{appendix}

\bibliographystyle{JHEP}
\bibliography{bibliography}

\providecommand{\href}[2]{#2}\begingroup\raggedright\begin{thebibliography}{10}

\bibitem{Benzke:2020htm}
M.~Benzke and T.~Hurth, \emph{{Resolved $1/m_b$ contributions to $\bar B \to X_{s,d} \ell^+\ell^-$ and $\bar B \to X_s \gamma$}}, \href{https://doi.org/10.1103/PhysRevD.102.114024}{\emph{Phys. Rev. D} {\bfseries 102} (2020) 114024}, [\href{https://arxiv.org/abs/2006.00624}{{\ttfamily 2006.00624}}].

\bibitem{Misiak:2026sqy}
M.~Misiak et~al., \emph{{The inclusive $\bar B \to X_s \gamma$ decay rate with higher precision}},  \href{https://arxiv.org/abs/2608.29863}{{\ttfamily 2608.29863}}.

\bibitem{Gunawardana:2019gep}
A.~Gunawardana and G.~Paz, \emph{{Reevaluating uncertainties in $\overline{B} \to X_{s}\gamma$ decay}}, \href{https://doi.org/10.1007/JHEP11(2019)141}{\emph{JHEP} {\bfseries 11} (2019) 141}, [\href{https://arxiv.org/abs/1908.02812}{{\ttfamily 1908.02812}}].

\bibitem{Bartocci:2024bbf}
R.~Bartocci, P.~B\"oer and T.~Hurth, \emph{{Renormalisation group evolution of the shape function g$_{17}$ in $\overline{B}\to {X}_s\gamma$ and $ \overline{B}\to {X}_s{\ell}^{+}{\ell}^{-} $ at subleading power}}, \href{https://doi.org/10.1007/JHEP04(2025)066}{\emph{JHEP} {\bfseries 04} (2025) 066}, [\href{https://arxiv.org/abs/2411.16634}{{\ttfamily 2411.16634}}].

\bibitem{Bartocci:2025lbl}
R.~Bartocci, P.~B{\"o}er and T.~Hurth, \emph{{NLO analysis of the subleading-power $Q_1$--$Q_{7\gamma}$ interference in $\bar{B} \to X_s \gamma$ at large photon energies}}, \href{https://doi.org/10.22323/1.485.0301}{\emph{PoS} {\bfseries EPS-HEP2025} (2026) 301}, [\href{https://arxiv.org/abs/2510.18811}{{\ttfamily 2510.18811}}].

\bibitem{Bartocci:2026eeq}
R.~Bartocci and T.~Hurth, \emph{Factorisation theorem and pole cancellation in the resolved {$Q_1^q$}-{$Q_{7\gamma}$} contribution to {$\bar{B}\to X_s\gamma$} at {NLO}},  \href{https://arxiv.org/abs/2610.14211}{{\ttfamily 2610.14211}}.

\bibitem{Bauer:2000ew}
C.~W. Bauer, S.~Fleming and M.~E. Luke, \emph{{Summing Sudakov logarithms in $B \to X_s \gamma$ in effective field theory}}, \href{https://doi.org/10.1103/PhysRevD.63.014006}{\emph{Phys. Rev. D} {\bfseries 63} (2000) 014006}, [\href{https://arxiv.org/abs/hep-ph/0005275}{{\ttfamily hep-ph/0005275}}].

\bibitem{Bauer:2000yr}
C.~W. Bauer, S.~Fleming, D.~Pirjol and I.~W. Stewart, \emph{{An Effective field theory for collinear and soft gluons: Heavy to light decays}}, \href{https://doi.org/10.1103/PhysRevD.63.114020}{\emph{Phys. Rev. D} {\bfseries 63} (2001) 114020}, [\href{https://arxiv.org/abs/hep-ph/0011336}{{\ttfamily hep-ph/0011336}}].

\bibitem{Beneke:2002ph}
M.~Beneke, A.~P. Chapovsky, M.~Diehl and T.~Feldmann, \emph{{Soft collinear effective theory and heavy to light currents beyond leading power}}, \href{https://doi.org/10.1016/S0550-3213(02)00687-9}{\emph{Nucl. Phys. B} {\bfseries 643} (2002) 431--476}, [\href{https://arxiv.org/abs/hep-ph/0206152}{{\ttfamily hep-ph/0206152}}].

\bibitem{Benzke:2010js}
M.~Benzke, S.~J. Lee, M.~Neubert and G.~Paz, \emph{{Factorization at Subleading Power and Irreducible Uncertainties in $\bar B\to X_s\gamma$ Decay}}, \href{https://doi.org/10.1007/JHEP08(2010)099}{\emph{JHEP} {\bfseries 08} (2010) 099}, [\href{https://arxiv.org/abs/1003.5012}{{\ttfamily 1003.5012}}].

\bibitem{Bosch:2004th}
S.~W. Bosch, B.~O. Lange, M.~Neubert and G.~Paz, \emph{{Factorization and shape function effects in inclusive B meson decays}}, \href{https://doi.org/10.1016/j.nuclphysb.2004.07.041}{\emph{Nucl. Phys. B} {\bfseries 699} (2004) 335--386}, [\href{https://arxiv.org/abs/hep-ph/0402094}{{\ttfamily hep-ph/0402094}}].

\bibitem{Becher:2006qw}
T.~Becher and M.~Neubert, \emph{{Toward a NNLO calculation of the $\bar{B} \to X_s \gamma$ decay rate with a cut on photon energy. II. Two-loop result for the jet function}}, \href{https://doi.org/10.1016/j.physletb.2006.04.046}{\emph{Phys. Lett. B} {\bfseries 637} (2006) 251--259}, [\href{https://arxiv.org/abs/hep-ph/0603140}{{\ttfamily hep-ph/0603140}}].

\bibitem{Bruser:2018rad}
R.~Br{\"u}ser, Z.~L. Liu and M.~Stahlhofen, \emph{{Three-Loop Quark Jet Function}}, \href{https://doi.org/10.1103/PhysRevLett.121.072003}{\emph{Phys. Rev. Lett.} {\bfseries 121} (2018) 072003}, [\href{https://arxiv.org/abs/1804.09722}{{\ttfamily 1804.09722}}].

\bibitem{Beneke:2020vnb}
M.~Beneke, P.~B\"oer, J.-N. Toelstede and K.~K. Vos, \emph{{QED factorization of non-leptonic $B$ decays}}, \href{https://doi.org/10.1007/JHEP11(2020)081}{\emph{JHEP} {\bfseries 11} (2020) 081}, [\href{https://arxiv.org/abs/2008.10615}{{\ttfamily 2008.10615}}].

\bibitem{Neubert:2004dd}
M.~Neubert, \emph{{Renormalization-group improved calculation of the $\bar{B} \to X_s \gamma$ branching ratio}}, \href{https://doi.org/10.1140/epjc/s2005-02141-1}{\emph{Eur. Phys. J. C} {\bfseries 40} (2005) 165--186}, [\href{https://arxiv.org/abs/hep-ph/0408179}{{\ttfamily hep-ph/0408179}}].

\bibitem{Greub:1996tg}
C.~Greub, T.~Hurth and D.~Wyler, \emph{{Virtual $\mathcal{O}(\alpha_s)$ corrections to the inclusive decay $b \to s \gamma$}}, \href{https://doi.org/10.1103/PhysRevD.54.3350}{\emph{Phys. Rev. D} {\bfseries 54} (1996) 3350--3364}, [\href{https://arxiv.org/abs/hep-ph/9603404}{{\ttfamily hep-ph/9603404}}].

\bibitem{Becher:2006nr}
T.~Becher and M.~Neubert, \emph{{Threshold resummation in momentum space from effective field theory}}, \href{https://doi.org/10.1103/PhysRevLett.97.082001}{\emph{Phys. Rev. Lett.} {\bfseries 97} (2006) 082001}, [\href{https://arxiv.org/abs/hep-ph/0605050}{{\ttfamily hep-ph/0605050}}].

\bibitem{Lepage:1979zb}
G.~P. Lepage and S.~J. Brodsky, \emph{{Exclusive Processes in Quantum Chromodynamics: Evolution Equations for Hadronic Wave Functions and the Form-Factors of Mesons}}, \href{https://doi.org/10.1016/0370-2693(79)90554-9}{\emph{Phys. Lett. B} {\bfseries 87} (1979) 359--365}.

\bibitem{Lepage:1980fj}
G.~P. Lepage and S.~J. Brodsky, \emph{{Exclusive Processes in Perturbative Quantum Chromodynamics}}, \href{https://doi.org/10.1103/PhysRevD.22.2157}{\emph{Phys. Rev. D} {\bfseries 22} (1980) 2157}.

\bibitem{Efremov:1979qk}
A.~V. Efremov and A.~V. Radyushkin, \emph{{Factorization and Asymptotical Behavior of Pion Form-Factor in QCD}}, \href{https://doi.org/10.1016/0370-2693(80)90869-2}{\emph{Phys. Lett. B} {\bfseries 94} (1980) 245--250}.

\bibitem{Beneke:2021pkl}
M.~Beneke, P.~B{\"o}er, J.-N. Toelstede and K.~K. Vos, \emph{{Light-cone distribution amplitudes of light mesons with QED effects}}, \href{https://doi.org/10.1007/JHEP11(2021)059}{\emph{JHEP} {\bfseries 11} (2021) 059}, [\href{https://arxiv.org/abs/2108.05589}{{\ttfamily 2108.05589}}].

\bibitem{Beneke:2009ek}
M.~Beneke, T.~Huber and X.-Q. Li, \emph{{NNLO vertex corrections to non-leptonic B decays: Tree amplitudes}}, \href{https://doi.org/10.1016/j.nuclphysb.2010.02.002}{\emph{Nucl. Phys. B} {\bfseries 832} (2010) 109--151}, [\href{https://arxiv.org/abs/0911.3655}{{\ttfamily 0911.3655}}].

\bibitem{Becher:2006mr}
T.~Becher, M.~Neubert and B.~D. Pecjak, \emph{{Factorization and Momentum-Space Resummation in Deep-Inelastic Scattering}}, \href{https://doi.org/10.1088/1126-6708/2007/01/076}{\emph{JHEP} {\bfseries 01} (2007) 076}, [\href{https://arxiv.org/abs/hep-ph/0607228}{{\ttfamily hep-ph/0607228}}].

\bibitem{Beneke:2001ev}
M.~Beneke, G.~Buchalla, M.~Neubert and C.~T. Sachrajda, \emph{{QCD factorization in $B\to \pi K, \pi\pi$ decays and extraction of Wolfenstein parameters}}, \href{https://doi.org/10.1016/S0550-3213(01)00251-6}{\emph{Nucl. Phys. B} {\bfseries 606} (2001) 245--321}, [\href{https://arxiv.org/abs/hep-ph/0104110}{{\ttfamily hep-ph/0104110}}].

\bibitem{Beneke:2002ni}
M.~Beneke and T.~Feldmann, \emph{{Multipole expanded soft collinear effective theory with nonAbelian gauge symmetry}}, \href{https://doi.org/10.1016/S0370-2693(02)03204-5}{\emph{Phys. Lett. B} {\bfseries 553} (2003) 267--276}, [\href{https://arxiv.org/abs/hep-ph/0211358}{{\ttfamily hep-ph/0211358}}].

\bibitem{Boer:2023yde}
P.~B{\"o}er and P.~Hager, \emph{{On the gauge-invariance of SCET beyond leading power}}, \href{https://doi.org/10.1007/JHEP08(2023)197}{\emph{JHEP} {\bfseries 08} (2023) 197}, [\href{https://arxiv.org/abs/2306.12412}{{\ttfamily 2306.12412}}].

\bibitem{Bodwin:2023asf}
G.~T. Bodwin, J.-H. Ee, D.~Kang and X.-P. Wang, \emph{{Gauge invariance of radiative jet functions in the position-space formulation of SCET}}, \href{https://doi.org/10.1103/PhysRevD.109.056020}{\emph{Phys. Rev. D} {\bfseries 109} (2024) 056020}, [\href{https://arxiv.org/abs/2302.05856}{{\ttfamily 2302.05856}}].

\bibitem{Beneke:1997zp}
M.~Beneke and V.~A. Smirnov, \emph{{Asymptotic expansion of Feynman integrals near threshold}}, \href{https://doi.org/10.1016/S0550-3213(98)00138-2}{\emph{Nucl. Phys. B} {\bfseries 522} (1998) 321--344}, [\href{https://arxiv.org/abs/hep-ph/9711391}{{\ttfamily hep-ph/9711391}}].

\bibitem{Shtabovenko:2016sxi}
V.~Shtabovenko, R.~Mertig and F.~Orellana, \emph{{New Developments in FeynCalc 9.0}}, \href{https://doi.org/10.1016/j.cpc.2016.06.008}{\emph{Comput. Phys. Commun.} {\bfseries 207} (2016) 432--444}, [\href{https://arxiv.org/abs/1601.01167}{{\ttfamily 1601.01167}}].

\bibitem{Shtabovenko:2020gxv}
V.~Shtabovenko, R.~Mertig and F.~Orellana, \emph{{FeynCalc 9.3: New features and improvements}}, \href{https://doi.org/10.1016/j.cpc.2020.107478}{\emph{Comput. Phys. Commun.} {\bfseries 256} (2020) 107478}, [\href{https://arxiv.org/abs/2001.04407}{{\ttfamily 2001.04407}}].

\bibitem{Shtabovenko:2023idz}
V.~Shtabovenko, R.~Mertig and F.~Orellana, \emph{{FeynCalc 10: Do multiloop integrals dream of computer codes?}}, \href{https://doi.org/10.1016/j.cpc.2024.109357}{\emph{Comput. Phys. Commun.} {\bfseries 306} (2025) 109357}, [\href{https://arxiv.org/abs/2312.14089}{{\ttfamily 2312.14089}}].

\bibitem{Mertig:1990an}
R.~Mertig, M.~Bohm and A.~Denner, \emph{{FEYN CALC: Computer algebraic calculation of Feynman amplitudes}}, \href{https://doi.org/10.1016/0010-4655(91)90130-D}{\emph{Comput. Phys. Commun.} {\bfseries 64} (1991) 345--359}.

\bibitem{Smirnov:2019qkx}
A.~V. Smirnov and F.~S. Chukharev, \emph{{FIRE6: Feynman Integral REduction with modular arithmetic}}, \href{https://doi.org/10.1016/j.cpc.2019.106877}{\emph{Comput. Phys. Commun.} {\bfseries 247} (2020) 106877}, [\href{https://arxiv.org/abs/1901.07808}{{\ttfamily 1901.07808}}].

\bibitem{Lee:2012cn}
R.~N. Lee, \emph{{Presenting LiteRed: a tool for the Loop InTEgrals REDuction}},  \href{https://arxiv.org/abs/1212.2685}{{\ttfamily 1212.2685}}.

\bibitem{Anastasiou:2002yz}
C.~Anastasiou and K.~Melnikov, \emph{{Higgs boson production at hadron colliders in NNLO QCD}}, \href{https://doi.org/10.1016/S0550-3213(02)00837-4}{\emph{Nucl. Phys. B} {\bfseries 646} (2002) 220--256}, [\href{https://arxiv.org/abs/hep-ph/0207004}{{\ttfamily hep-ph/0207004}}].

\bibitem{Kotikov:1990kg}
A.~V. Kotikov, \emph{{Differential equations method: New technique for massive Feynman diagrams calculation}}, \href{https://doi.org/10.1016/0370-2693(91)90413-K}{\emph{Phys. Lett. B} {\bfseries 254} (1991) 158--164}.

\bibitem{Kotikov:1991hm}
A.~V. Kotikov, \emph{{Differential equations method: The Calculation of vertex type Feynman diagrams}}, \href{https://doi.org/10.1016/0370-2693(91)90834-D}{\emph{Phys. Lett. B} {\bfseries 259} (1991) 314--322}.

\bibitem{Gehrmann:1999as}
T.~Gehrmann and E.~Remiddi, \emph{{Differential equations for two loop four point functions}}, \href{https://doi.org/10.1016/S0550-3213(00)00223-6}{\emph{Nucl. Phys. B} {\bfseries 580} (2000) 485--518}, [\href{https://arxiv.org/abs/hep-ph/9912329}{{\ttfamily hep-ph/9912329}}].

\bibitem{Henn:2013pwa}
J.~M. Henn, \emph{{Multiloop integrals in dimensional regularization made simple}}, {\emph{Phys. Rev. Lett.} {\bfseries 110} (2013) 251601}, [\href{https://arxiv.org/abs/1304.1806}{{\ttfamily 1304.1806}}].

\bibitem{Meyer:2017joq}
C.~Meyer, \emph{{Algorithmic transformation of multi-loop master integrals to a canonical basis with CANONICA}}, \href{https://doi.org/10.1016/j.cpc.2017.09.014}{\emph{Comput. Phys. Commun.} {\bfseries 222} (2018) 295--312}, [\href{https://arxiv.org/abs/1705.06252}{{\ttfamily 1705.06252}}].

\bibitem{Meyer:2016slj}
C.~Meyer, \emph{{Transforming differential equations of multi-loop Feynman integrals into canonical form}}, \href{https://doi.org/10.1007/JHEP04(2017)006}{\emph{JHEP} {\bfseries 04} (2017) 006}, [\href{https://arxiv.org/abs/1611.01087}{{\ttfamily 1611.01087}}].

\bibitem{Weinzierl:2022eaz}
S.~Weinzierl, \emph{{Feynman Integrals. A Comprehensive Treatment for Students and Researchers}}.
\newblock UNITEXT for Physics. Springer, 2022, \href{https://doi.org/10.1007/978-3-030-99558-4}{10.1007/978-3-030-99558-4}.

\bibitem{goncharov2011multiplepolylogarithmscyclotomymodular}
A.~B. Goncharov, \emph{{Multiple polylogarithms, cyclotomy and modular complexes}},  \href{https://arxiv.org/abs/1105.2076}{{\ttfamily 1105.2076}}.

\bibitem{3137480}
R.~Bartocci, \emph{Synergies between effective field theories: global analyses of the Minimal Flavour Violating SMEFT and subleading NLO analysis of $b\to s\gamma$ transitions}, Ph.D. thesis, Johannes Gutenberg-Universit\"at Mainz, 2026.

\bibitem{Lange:2003ff}
B.~O. Lange and M.~Neubert, \emph{{Renormalization group evolution of the B meson light cone distribution amplitude}}, \href{https://doi.org/10.1103/PhysRevLett.91.102001}{\emph{Phys. Rev. Lett.} {\bfseries 91} (2003) 102001}, [\href{https://arxiv.org/abs/hep-ph/0303082}{{\ttfamily hep-ph/0303082}}].

\bibitem{Beneke:2022msp}
M.~Beneke, P.~B{\"o}er, J.-N. Toelstede and K.~K. Vos, \emph{{Light-cone distribution amplitudes of heavy mesons with QED effects}}, \href{https://doi.org/10.1007/JHEP08(2022)020}{\emph{JHEP} {\bfseries 08} (2022) 020}, [\href{https://arxiv.org/abs/2204.09091}{{\ttfamily 2204.09091}}].

\bibitem{Schwinger:1960qe}
J.~S. Schwinger, \emph{{Brownian motion of a quantum oscillator}}, \href{https://doi.org/10.1063/1.1703727}{\emph{J. Math. Phys.} {\bfseries 2} (1961) 407--432}.

\bibitem{Keldysh:1964ud}
L.~V. Keldysh, \emph{{Diagram technique for nonequilibrium processes}}, {\emph{Zh. Eksp. Teor. Fiz.} {\bfseries 47} (1964) 1515--1527}.

\bibitem{Becher:2007ty}
T.~Becher, M.~Neubert and G.~Xu, \emph{{Dynamical Threshold Enhancement and Resummation in Drell-Yan Production}}, \href{https://doi.org/10.1088/1126-6708/2008/07/030}{\emph{JHEP} {\bfseries 07} (2008) 030}, [\href{https://arxiv.org/abs/0710.0680}{{\ttfamily 0710.0680}}].

\bibitem{Bell:2009nk}
G.~Bell, \emph{{NNLO vertex corrections in charmless hadronic B decays: Real part}}, \href{https://doi.org/10.1016/j.nuclphysb.2009.07.012}{\emph{Nucl. Phys. B} {\bfseries 822} (2009) 172--200}, [\href{https://arxiv.org/abs/0902.1915}{{\ttfamily 0902.1915}}].

\end{thebibliography}\endgroup

\end{document}